\documentclass[useAMS,usenatbib,twocolumn]{mn2e}
\pdfoutput=1
\usepackage{amssymb}
\usepackage{amsmath}
\usepackage{natbib}
\usepackage[pdftex]{graphicx}
\usepackage{subfigure}
\usepackage{booktabs}
\begin{document}

\title[Tidal instability]{Nonlinear evolution of the elliptical instability in the presence of weak magnetic fields}
   \author[A.J. Barker \& Y. Lithwick]{Adrian J. Barker\thanks{E-mail:
	  adrianjohnbarker@gmail.com} and Yoram Lithwick \\
	  Center for Interdisciplinary Exploration and Research in Astrophysics (CIERA) \& \\ Dept. of Physics and Astronomy, Northwestern University, 2145 Sheridan Rd, Evanston, IL 60208, USA.}
	
\pagerange{\pageref{firstpage}--\pageref{lastpage}} \pubyear{2013}

\maketitle

\label{firstpage}

\begin{abstract}
We investigate whether the elliptical instability is important for tidal dissipation in gaseous planets and stars. In a companion paper, we found that the conventional elliptical instability results in insufficient dissipation because it
 produces long-lived vortices that then quench further instability. Here, we study whether the addition of a magnetic field prevents those vortices from forming, and hence leads to enhanced dissipation. We present results from  magnetohydrodynamic simulations that evolve
 the elliptical instability in a local patch of a rotating planet or star, in the presence of a weak magnetic field.
 We find that magnetic fields do indeed prevent vortices from forming, and hence greatly enhance the steady state dissipation rate. In addition, the resulting turbulence acts as a small-scale dynamo, amplifying the initially weak field. The inferred tidal dissipation is potentially important at short orbital periods. For example, it can circularise hot Jupiters with orbital periods shorter than $2.5$ days, and synchronise their spins with their orbits out to $6$ days. However, it appears unable to account for the hot Jupiters that appear to have been circularised out to six to ten day orbital periods.
 It also cannot account for the inferred circularisation of many close binary stars.
\end{abstract}


\section{Introduction}

The orbits and spins of short-period gaseous extrasolar planets are thought to have been significantly modified by tidal evolution. Dissipation of the tidal flows in fluid regions of the planetary interior, excited by stellar gravitational forcing, is thought to be responsible for circularising an initially eccentric orbit, and synchronising the spins of these planets with their orbits. Unfortunately, the mechanisms of tidal dissipation are poorly understood, particularly for rotating fluid bodies, such as giant planets and stars.
 In a previous paper (BL13), we began an investigation into the nonlinear evolution of the elliptical instability, to determine whether the resulting turbulence was sufficiently dissipative to explain the inferred levels of tidal dissipation in fluid planets and stars. In this paper, we extend our investigation by introducing a weak magnetic field, and study how this modifies the dissipative properties of the turbulence. For a more detailed introduction to this problem see \cite{BL13}, hereafter referred to as BL13.

The response of a fluid body to tidal forcing is often decomposed into two components. One is the equilibrium tide, which is a quasi-hydrostatic spheroidal bulge that follows the motion of the companion. This produces elliptical streamlines in the flow inside a rotating fluid body. This component is usually thought to be dissipated through its interaction with turbulent convection
 \citep{Zahn1966}, though the efficiency of this process is still debated (e.g.~\citealt{Goldreich1977}; \citealt{Penev2007}; \citealt{OgilvieLesur2012}). The other component is called the dynamical tide, which represents the tidal excitation of internal waves, in particular those restored by buoyancy and/or rotation. The dissipation of these waves could also lead to tidal evolution (\citealt{Zahn1977,GoodmanDickson1998,Terquem1998,Gio2004,Wu2005b,Barker2010}).

However, the aforementioned dissipation mechanisms are generally too weak. For example, they do not appear to account for the hot Jupiters with circular orbits out to orbital periods of six to ten days\footnote{e.g.~http://www.openexoplanetcatalogue.com}, if these started out in eccentric orbits and were subsequently tidally circularised.
This motivates us to examine a less traditional mechanism:
 the equilibrium tide can be subject to parametric instabilities. These can excite waves restored by
  buoyancy forces \citep{Weinberg2012} or by the Coriolis
force---the latter case is called the elliptical instability (e.g.~\citealt{Kerswell2002}).
   The elliptical instability can be thought of as an instability of elliptical streamlines, such as those of the equilibrium tidal flow in a rotating fluid planet. It drives inertial waves through a parametric resonance. It is a nonlinear mechanism of tidal dissipation, and so is not represented in most previous studies, which typically make the assumption of linearity. 
 \cite{Rieutord2004} proposed that the elliptical instability might be important for tidal dissipation in close binary stars.
   The elliptical instability has been studied in both laboratory experiments \citep{Lacaze2004,LeBars2010} and numerical simulations \citep{Cebron2010,Cebron2012}.    
   However, the outcome of the instability has yet to be quantified in the astrophysical regime of small ellipticity. Most importantly, it remains
   to be seen whether the instability drives turbulence in steady state, and if so, whether the amount
   of turbulent dissipation is sufficient to be astrophysically important.

In BL13, we began an investigation into this problem, by performing hydrodynamical simulations of a small patch of a rotating tidally deformed planet or star. We neglected buoyancy forces, which is appropriate for studying an approximately adiabatically stratified convective region. We found that the instability saturates differently depending on the ellipticity of the streamlines, or equivalently, the ratio of strain to rotation in the elliptical flow. In the astrophysically relevant regime of small ellipticity, our simulations indicate that the instability saturates through the formation of coherent columnar (``Proudman-Taylor") vortices, whose presence inhibits the driving mechanism of the instability. This results in much weaker dissipation than would be predicted if the instability is continuously driven. The dissipation in the astrophysical regime was found to be negligible, suggesting that the elliptical instability is unlikely to be responsible for tidal dissipation in planets and stars. However, our numerical model neglected to include magnetic fields, which is the extra ingredient that we introduce in this work.

Jovian planets are likely to harbour and generate their own internal magnetic fields. Indeed, Jupiter is observed to contain a large-scale magnetic field at the surface, which is thought to be generated by dynamo action driven by buoyancy forces in the deep convective interior \citep{Glatz2010,Jones2011}. Short-period Jovian planets are also very likely to generate their own internal magnetic fields. The external manifestation of these fields might be detectable through radio emission caused by star-planet magnetic interaction, analogous to the Jupiter-Io plasma interaction \citep{Stevens2005,Zarka2007}. However, so far no observations have been successful in detecting any planetary radio emission, and so the strengths of their magnetic fields remain unconstrained.

Since the Jovian convective interior is magnetised, and this is also very likely to be true for the interiors of short-period gaseous extrasolar planets, it is essential to study the addition of magnetic fields on the nonlinear evolution of the elliptical instability. This is because even a weak magnetic field is able to fundamentally change the stability properties of a flow. For example, the addition of even a very weak magnetic field to the flow in a Keplerian accretion disc renders the fluid unstable to the magneto-rotational instability \citep{Balbus1991}. This motivates us to investigate in this paper whether a weak magnetic field can destroy or inhibit the formation of vortices, which might result in more efficient dissipation than was observed in our previous hydrodynamical simulations (BL13)\footnote{The properties of the elliptical instability are also modified in the presence of a sufficiently strong magnetic field \citep{Lebovitz2004,Mizerski2009,MizBaj2011}. But in this paper, the initial field is assumed to be sufficiently weak  that the initial instability
is essentially equivalent to the purely hydrodynamical elliptical instability.
Interestingly, the generalised magneto-elliptical instability corresponds with the magneto-rotational instability when the ellipticity of the streamlines is taken to be infinite \citep{MizLyr2012}.}.

In this paper we study the elliptical instability in the presence of a weak magnetic field. The structure of the paper is as follows. We first present 
an order of magnitude estimate for the dissipation resulting from the elliptical instability in \S \ref{potential}, together with our estimates of the relevance of magnetic fields. We then summarise our numerical model (explained in more detail in BL13) in \S \ref{local}. Our results are presented in \S 4-6, followed by a discussion and conclusion, where we apply the results of our investigation to astrophysical tidal dissipation.

\section{Potential astrophysical importance of the elliptical instability}
\label{potential}
In BL13, we estimated the potential astrophysical importance of the dissipation resulting from the elliptical instability. Our argument is briefly outlined here. 

We consider the elliptical tidal flow in a planet of mass $m_{p}$, induced by the tidal potential due to a star of mass $m_{\star}$, for the case in which the rotation of the planet $\Omega$ is not equal to the orbital mean motion of its orbit around the star $n$. For simplicity,  we assume that the planet is on a circular orbit in the equatorial plane of the primary. The fastest growing mode of the elliptical instability has a growth rate approximately $\epsilon \gamma/2$ (e.g.~\citealt{Kerswell2002}) where 
\begin{eqnarray}
\gamma=\Omega-n,
\end{eqnarray}
is half the tidal frequency, and $\epsilon$ is the ellipticity, set by the tidal potential of the star. Note that if the radial displacement in the equilibrium tide is $\xi_{r}$, and the radius of the planet is $R_{p}$, then
\begin{eqnarray}
\epsilon=\frac{\xi_{r}}{R_{p}}\approx\frac{m_{\star}}{m_{p}+m_{\star}}\left(\frac{P_{dyn}}{P}\right)^{2}\sim 10^{-2}\left(\frac{1 \; \mathrm{d}}{P}\right)^{2}, 
\end{eqnarray}
where $P=2\pi/n$ is the orbital period and is approximately one day for the shortest-period hot Jupiters and $P_{dyn}=2\pi \sqrt{\frac{R_{p}^{3}}{G m_{p}}}$, is the dynamical timescale.

We hypothesise that the nonlinear outcome of the elliptical instability is a turbulent steady state in which the growth rate of the modes that dominate the energy is balanced by their nonlinear cascade rate. Then $\epsilon |\gamma| \sim u/l$, where the energy-dominating modes have spatial scale $l$, and velocity amplitude $u$. The resulting dissipation rate, assuming sustained energy injection, is $D \sim m_{p}u^{3}/l \sim m_{p}l^{2} \epsilon^{3}|\gamma|^{3}$.  We therefore introduce the dimensionless efficiency factor $\chi$, by
\begin{eqnarray}
D =\chi m_{p} (2R_{p})^{2} |\gamma|^{3} \epsilon^{3}.
\end{eqnarray}
If $\chi \sim 1$, this mechanism can explain the circular orbits of hot Jupiters out to approximately $4.5$ days, and similarly, it predicts the synchronisation of their spins with their orbits out to $11.5$ days (BL13 and \S \ref{discussion} below). These estimates suggest that the elliptical instability could play an important role in contributing to the orbital circularisation of initially eccentric hot Jupiters, and could potentially provide an explanation for the observed preponderance of circular orbits amongst the shortest period planets, which has so far eluded explanation. However, it may not be appropriate to take $\chi\sim 1$. In our previous hydrodynamical simulations (BL13), the instability was prevented from depositing energy at an efficient rate once columnar vortices had formed. When this occurs, $\chi \ll 1$, at least in the astrophysical regime of small ellipticity. 

In this paper we study the dissipative properties of the turbulence driven by the elliptical instability, in the presence of a weak magnetic field, in order to quantify the efficiency of the dissipation, $\chi$, and its relevance for astrophysical observations.

\subsection{Presence of magnetic fields}
\label{magneticestimates}

A simple estimate for the strength of the magnetic field that may be relevant for this problem is as follows. If the elliptical instability generates flows with velocity amplitude $u\sim \epsilon \gamma R_{p}$, then the associated kinetic energy density is $K\sim \frac{1}{2}\bar{\rho}\epsilon^{2}\gamma^{2}R_{p}^{2}$, where $\bar{\rho}$ is the mean density. This can be compared with the magnetic energy density 
$M \sim B^{2}/8\pi$. If we take $\gamma \sim 2\pi/P$, where $P$ is the orbital period, then, for a short-period hot Jupiter,
\begin{eqnarray}
\nonumber
\frac{M}{K} &\approx& 2 \times 10^{-7} \left(\frac{1 \mathrm{g cm^{-3}}}{\bar{\rho}}\right)\left(\frac{B}{10 G}\right)^{2}\left(\frac{R_{J}}{R}\right)^{2} \\
&& \hspace{2.5cm} \times \left(\frac{2.8\mathrm{hr}}{P_{dyn}}\right)^{4}\left(\frac{P}{1\mathrm{d}}\right)^{6},
\end{eqnarray}
where we have scaled the magnetic field to that inferred for Jupiter's dipolar field at the surface \citep{Jones2011}.
Therefore for a non-synchronised hot Jupiter, the magnetic field is likely to be very weak in comparison with the kinetic energy of the flows generated by the elliptical instability. However, as we have discussed in the introduction, it is known that even a very weak magnetic field can drastically alter the stability properties of a flow. The inclusion of such a magnetic field may change the properties of the turbulence generated by the elliptical instability, which may lead to enhanced tidal dissipation, over cases without a magnetic field.

\section{Local model of a tidally deformed body}
\label{local}

Our model considers a small patch of a spinning hot Jupiter on a circular orbit about its star, focusing on the synchronisation problem\footnote{Extensions 
to other situations, e.g., tidal circularisation and tides in binary stars, are discussed in \S \ref{discussion}. \label{footnote:other}}.
We neglect buoyancy forces, as is appropriate in an adiabatically stratified convection zone, and neglect the additional influence of turbulent convection.
The spin and orbital angular velocity vectors are $\boldsymbol{\Omega}=\Omega \boldsymbol{e}_{z}$ and $n \boldsymbol{e}_{z}$.
In the rotating frame of the orbit (frequency $n$), we model 
  the equilibrium tide
 as an elliptical flow in the hot Jupiter: each fluid element 
traces an ellipse  that is fixed in this frame (and points at the star), and 
the period to trace out the ellipse is
  $2\pi/|\gamma|$ (where $\gamma=\Omega-n$).
  Both this period and the ellipticity of the ellipses ($\epsilon$) are taken to be independent
  of distance from the planet's center.
      For our numerical calculations, we work in the frame of the planet's
 spin ($\Omega$) rather than its orbit ($n$), because in that case the tidal perturbation
 is  a small  ($O(\epsilon)$) correction to flow in a rotating frame.

 The background flow in the frame of the planet's spin is
\begin{eqnarray}
\boldsymbol{U}_{0} = \mathrm{A}\boldsymbol{x} =-\gamma \epsilon\left(\begin{array}{ccc}
s_{2\gamma t} &  c_{2\gamma t} & 0 \\
c_{2\gamma t} & -s_{2\gamma t} & 0 \\
0 & 0 & 0 \end{array}\right)\boldsymbol{x},
\label{eq:A}
\end{eqnarray}
where $\cos 2\gamma t \equiv c_{2\gamma t}$ and  $\sin 2\gamma t \equiv s_{2\gamma t}$ (e.g.~BL13).
We assume that this flow is perfectly maintained, so there is an infinite reservoir of energy to drive the instability. Our aim is to determine the dissipative properties of the turbulence produced by the instability of this background flow. For further details of our model, see BL13.

We consider an incompressible three-dimensional rotating fluid, satisfying the Navier-Stokes equations for the evolution of perturbations to the base flow $\boldsymbol{U}_{0}$. Including the induction equation for the magnetic field, we have
\begin{eqnarray}
\label{eq1a}
&& D\boldsymbol{u}=-\nabla p -  2\boldsymbol{\Omega}\times\boldsymbol{u} -\mathrm{A}\boldsymbol{u} + \boldsymbol{B}\cdot \nabla\boldsymbol{B} +D^{\nu}_{\alpha}(\boldsymbol{u}), \\
&& D\boldsymbol{B}= \boldsymbol{B}\cdot \nabla\boldsymbol{u} + \mathrm{A}\boldsymbol{B} +D^{\eta}_{\alpha}(\boldsymbol{B}), \\
&& \nabla \cdot \boldsymbol{u} = 0, \\
&& \nabla \cdot \boldsymbol{B} = 0, \\
&& D \equiv \partial_{t}+ \boldsymbol{U}_{0}\cdot \nabla + \boldsymbol{u}\cdot \nabla,
\label{eq1b}
\end{eqnarray}
subject to appropriate boundary conditions, where $\boldsymbol{u}$ is the velocity, $\boldsymbol{B}$ is the magnetic field, and the modified pressure $p$ contains a contribution from the magnetic pressure. Our unit of length is the size of the box $L$, time is in units of $\gamma^{-1}$, the density $\rho=1$, and $\boldsymbol{B}$ is normalised by the Alfv\'{e}n speed. We define the dissipative operators $D^{\nu}_{\alpha}(\boldsymbol{u})= \nu_{\alpha} (-1)^{\alpha+1}\nabla^{2\alpha}\boldsymbol{u}$ and $D^{\eta}_{\alpha}(\boldsymbol{B})= \eta_{\alpha} (-1)^{\alpha+1}\nabla^{2\alpha}\boldsymbol{B}$, which reduce to the standard viscous and ohmic diffusion operators when $\alpha=1$,

The mean kinetic ($K=\frac{1}{2}\langle |\boldsymbol{u}|^{2}\rangle$), magnetic ($M=\frac{1}{2}\langle |\boldsymbol{B}|^{2}\rangle$) and total energies ($E=K+M$) evolve according to,
\begin{eqnarray}
\partial_{t}K &=& -\langle\boldsymbol{u}\mathrm{A}\boldsymbol{u}\rangle + \langle \boldsymbol{u}\cdot \left(\boldsymbol{B}\cdot \nabla\boldsymbol{B}\right)\rangle-D_{K}, \\ 
\partial_{t}M &=& \langle\boldsymbol{B}\mathrm{A}\boldsymbol{B}\rangle - \langle \boldsymbol{u}\cdot \left(\boldsymbol{B}\cdot \nabla\boldsymbol{B}\right)\rangle-D_{M}, \\ 
\partial_{t}E &=& -\langle\boldsymbol{u}\mathrm{A}\boldsymbol{u}\rangle + \langle\boldsymbol{B}\mathrm{A}\boldsymbol{B}\rangle -D_{K}-D_{M},
\end{eqnarray}
where
\begin{eqnarray}
D_{K} &=& \nu_{\alpha} (-1)^{\alpha+1}\langle \boldsymbol{u}\cdot\nabla^{2\alpha}\boldsymbol{u}\rangle, \\
D_{M} &=& \eta_{\alpha} (-1)^{\alpha+1}\langle \boldsymbol{B}\cdot\nabla^{2\alpha}\boldsymbol{B}\rangle,
\end{eqnarray}
which, when $\alpha=1$, reduce to $D_{K} = \nu \langle |\boldsymbol{\omega}|^{2}\rangle$ and $D_{M} = \eta \langle |\boldsymbol{J}|^{2}\rangle$, where the vorticity is $\boldsymbol{\omega}=\nabla\times\boldsymbol{u}$, and the current density is $\boldsymbol{J}=\nabla\times\boldsymbol{B}$. We have defined the averaging operation $\langle \chi \rangle =  \frac{1}{L^{3}}\int_{V} \chi dV$.
For this problem the most important quantity is the total volumetric dissipation rate, $D = D_{K} +D_{M}$. 

We expand perturbations into a set of shearing waves \citep{Kelvin1880}, which are plane waves with evolving wavevectors $\boldsymbol{k}(t)$, such that:
\begin{eqnarray}
\nonumber
\left[\boldsymbol{u}(\boldsymbol{x},t),p(\boldsymbol{x},t),\boldsymbol{B}(\boldsymbol{x},t)\right]= \mathrm{Re}\left\{\left[\hat{\boldsymbol{u}}(t),\hat{p}(t),\hat{\boldsymbol{B}}(t)\right] e^{i \boldsymbol{k}(t)}\right\}.
\end{eqnarray}
The wavevector evolves according to
\begin{eqnarray}
\label{kvol1}
\dot{\boldsymbol{k}} + \mathrm{A}^{T}\boldsymbol{k} = 0.
\end{eqnarray}
This is similar to the shearing box in an accretion disc (e.g.~\citealt{GoldreichLyndenBell1965}; \citealt{Hawley1995}). Such a method allows the use of a Cartesian pseudo-spectral code, for which we use SNOOPY (\citealt{Lesur2005}; \citealt{Lesur2007}). We have modified the code so that it solves Eqs.~\ref{eq1a}--\ref{eq1b} by expanding the flow in terms of shearing waves that satisfy Eq.~\ref{kvol1}. This has been tested for the hydrodynamical problem in BL13. We have also verified that the magnetic field evolves correctly according to the linearised induction equation for some initial trial magnetic fields, by comparing the solutions with Mathematica. 

We impose an initial magnetic field, of the form
\begin{eqnarray}
\label{B0}
\boldsymbol{B}(\boldsymbol{x},t=0) = B_{0}\sin \left(\frac{2\pi x}{L}\right)\boldsymbol{e}_{z},
\end{eqnarray}
which has $\langle \boldsymbol{B} \rangle=0$, implying zero net flux. This form of the magnetic field was chosen so that the energy source that drives any instabilities is ultimately due to the background elliptical flow. Given the estimates in \S \ref{magneticestimates}, we consider $|B_{0}| \ll \epsilon$, usually taking $\epsilon\sim [10^{-2},0.2]$ and $B_{0}\in [10^{-3},10^{-2}]$. The magnetic energy is therefore initially weak in comparison with the tidal flow by the factor $B_{0}^{2}/\epsilon^{2} =[2.5\times 10^{-5},10^{-2}]$. Our initial conditions for the velocity field consist of solenoidal large scale noise (with amplitude $\sim0.01$) for wavenumbers
 $\frac{k}{2\pi}\in[1,12]$.

The interiors of giant planets ohmically diffuse magnetic fields faster than they diffuse momentum by viscosity, i.e., the magnetic Prandtl number Pm $\equiv\nu/\eta \ll 1$. For the interior of Jupiter it is estimated that, depending on location within the convective regions of the planet, $10^{-6}  \lesssim \mathrm{Pm} \lesssim 10^{-2}$, whereas for a stellar interior, values of Pm as low as $10^{-8}$ can be expected \citep{Fauve2005,Jones2011}. In our simulations, this is a very difficult parameter regime to reach, so we begin by adopting $\nu=\eta$. We later check the dependence on this parameter in \S \ref{Pm} by reaching small ``effective" magnetic Prandtl numbers using hyperdiffusion. We primarily use hyperdiffusion because this allows us to study a wider range in $\epsilon$ than standard viscosity, though simulations with standard viscosity have also been performed, and will be presented in \S \ref{Standard}.

To explore the parameter space, most of the simulations in this paper use a modest resolution of $128^3$. This is for two reasons: firstly we wish to run them for sufficient time for meaningful averages to be computed from the turbulent state, and it is feasible to run $128^3$ much longer than at higher resolution. Secondly, we wish to survey the parameter space to study how $D$ varies with $\epsilon,\Omega,\nu,\eta,\alpha,\boldsymbol{B}_{0}$, and this is only computationally feasible with simulations at this resolution or lower. However, several simulations at higher resolution have been performed, as outlined in Table \ref{table}, and as described in the text. 

\label{Results}
\section{Illustrative simulation}
\label{illustrative}

\begin{figure}
  \begin{center}
               \subfigure{\includegraphics[trim=0cm 0cm 0cm 0cm, clip=true,width=0.55\textwidth]{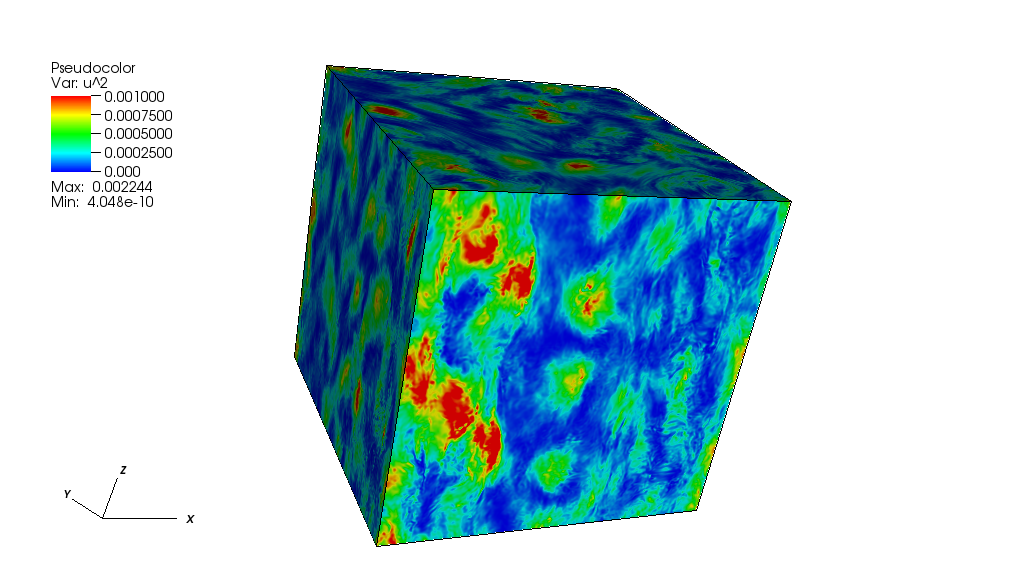} }	
              \subfigure{\includegraphics[trim=0cm 0cm 0cm 0cm, clip=true,width=0.55\textwidth]{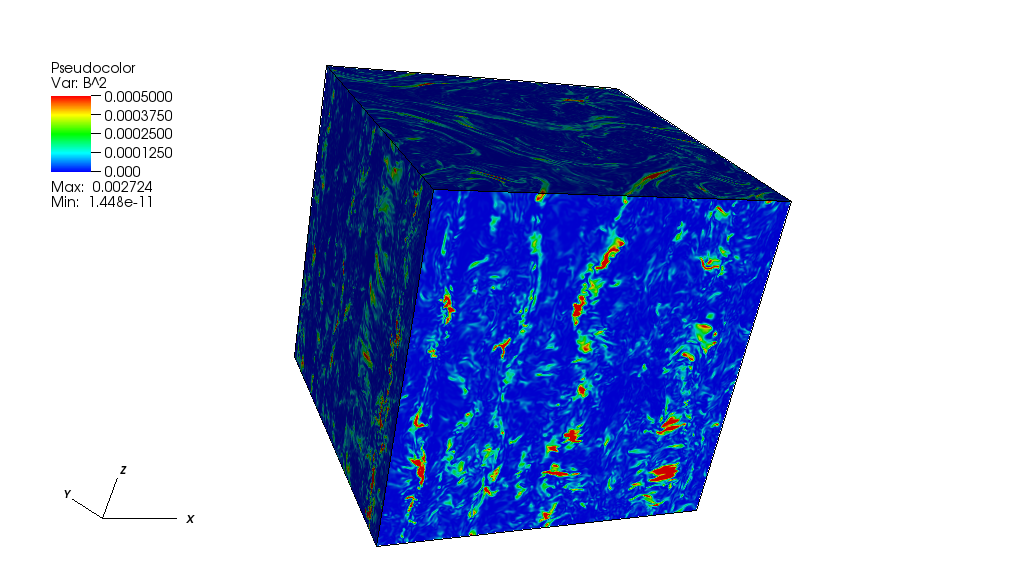} }	
               \end{center}
  \caption{Visualisation of $|\boldsymbol{u}|^{2}$ and $|\boldsymbol{B}|^{2}$ in our highest resolution simulation with $\epsilon=0.1$ and $n=0$, simulated with $512^3$ gridpoints and $\alpha=4$ and $\nu_{4}=\eta_{4}=10^{-23}$ at $t=300$.}
  \label{5}
\end{figure}
 \begin{figure}
  \begin{center}
      \subfigure{\includegraphics[trim=0cm 0cm 0cm 0cm, clip=true,width=0.45\textwidth] {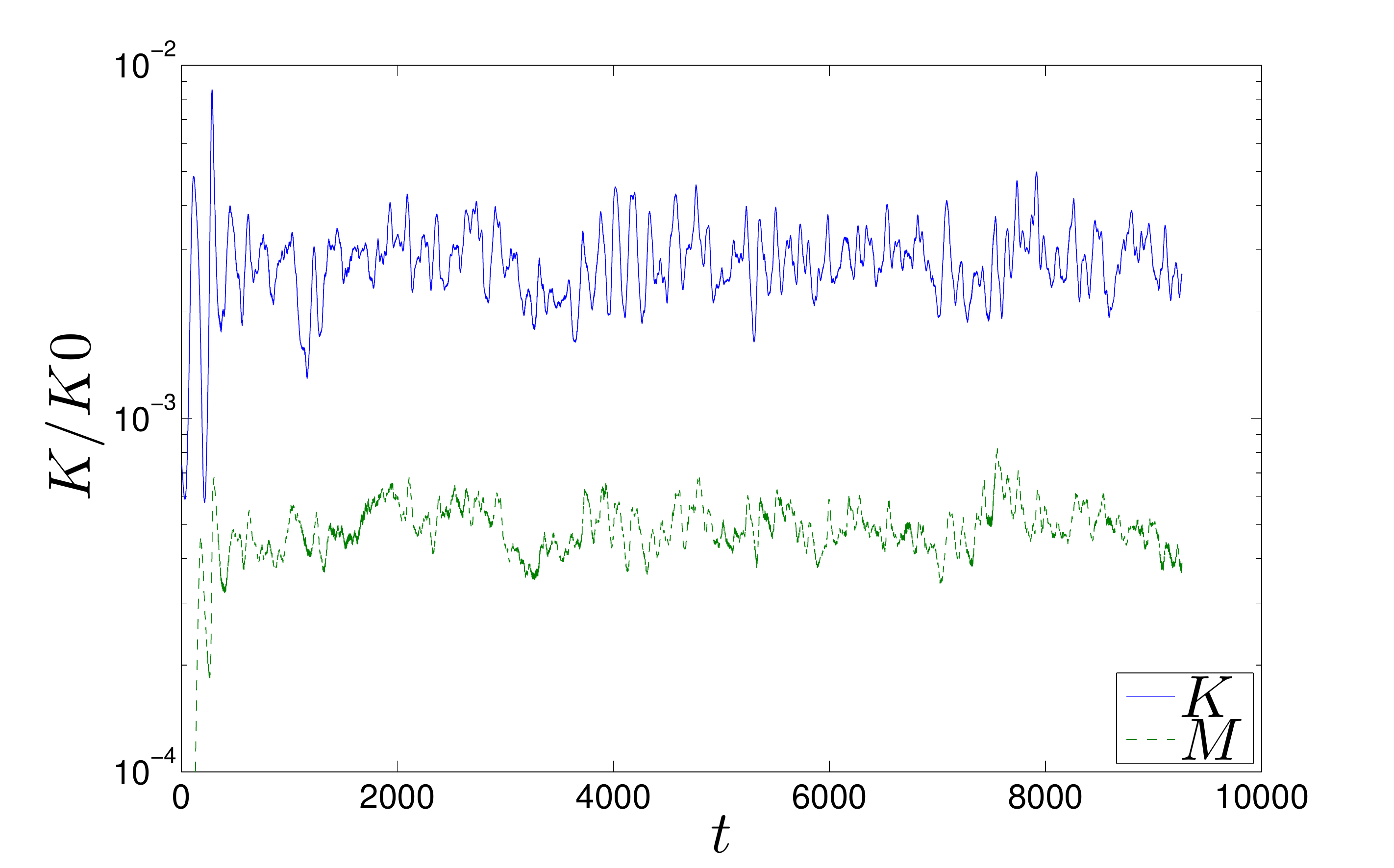} }  \\ 
          \subfigure{\includegraphics[trim=0cm 0cm 0cm 0cm, clip=true,width=0.45\textwidth] {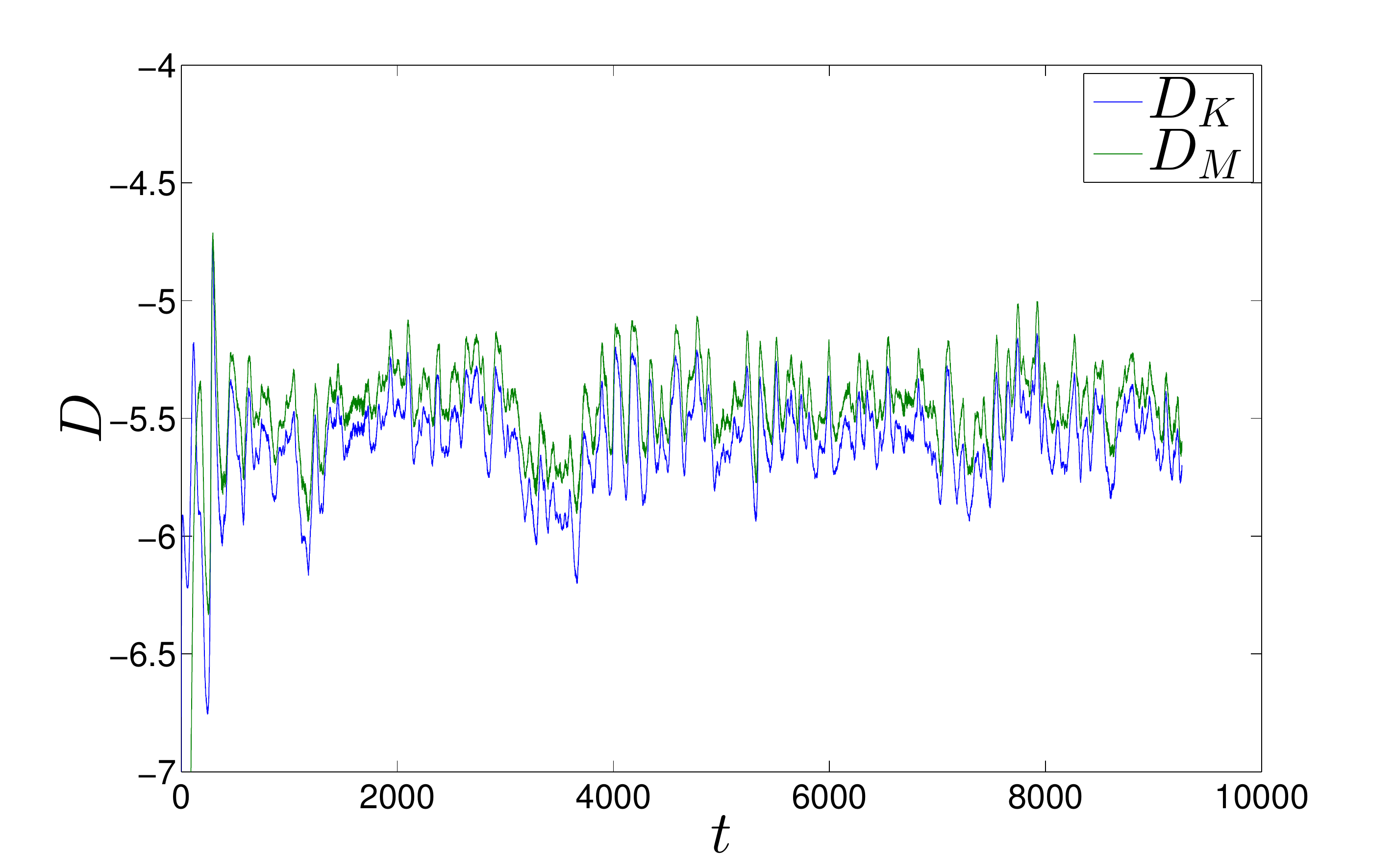} }  \\
               \end{center}
  \caption{Volume integrated quantities for our illustrative simulation with $n=0$, $\epsilon=0.1$, $\alpha=4$ and $\nu_{4}=\eta_{4}=10^{-18}$. Top: kinetic and magnetic energies normalised to the background vortex. Bottom: viscous and ohmic dissipation.}
  \label{1}
\end{figure}

For an illustrative simulation, we adopt a stationary bulge, with $n=0$ (i.e.~$\Omega=1$), and $\epsilon=0.1$, in
a unit box, using hyperdiffusion for both the viscous and ohmic diffusion operators ($\alpha=4$) with $\nu_{4}=\eta_{4}=10^{-18}$, and start with an initial magnetic field with magnitude $B_{0}=10^{-3}$. Comparison simulations performed with standard ohmic and viscous diffusion operators, as well as a study of the dependence of our results on $\nu_{\alpha},\eta_{\alpha}$ are presented in \S \ref{Standard}.

Since the magnetic field is initially weak, the initial evolution is very similar to the hydrodynamical simulations described in BL13. The magnetic field is not sufficiently strong to modify the linear instability i.e., the waves that are excited are inertial waves, only very weakly affected by the magnetic field. These waves grow until they reach sufficient amplitude for nonlinearities to limit their growth, leading to turbulence. During this initial turbulent phase, nonlinear interactions between inertial waves transfer energy into the modes with $k_{z}=0$ and the flow becomes rapidly dominated by columnar vortices aligned with the rotation axis. However, in the presence of a weak magnetic field, magnetic stresses destroy these vortices, leading to sustained turbulence, which amplifies the initially weak magnetic field.  An illustrative visualisation of $|\boldsymbol{u}|^{2}$ and $|\boldsymbol{B}|^{2}$ in a snapshot in the turbulent state at $t=300$ is shown (from our highest resolution simulation with $512^3$) in Fig.~\ref{5}. The flow remains turbulent in steady state. By contrast, the hydrodynamical simulations in BL13 were dominated by long-lived columnar vortices.

Volume-integrated quantities from this simulation are plotted in Fig.~\ref{1}. In the top panel, we show the temporal evolution of the kinetic and magnetic energies (both normalised to the kinetic energy in the background vortex, $\frac{1}{12}\left(\Omega^{2}+(\gamma\epsilon)^{2}\right)$). In the bottom panel we plot the viscous and ohmic dissipation rates as a function of time. The total dissipation has both viscous and ohmic contributions, each having similar magnitudes, though the latter is slightly larger.

Inertial waves are continually driven at large scales, and the energy is cascaded to small scales and ultimately dissipated by a combination of viscous and ohmic diffusion. Hence, the elliptical instability, in the presence of a weak magnetic field, leads to sustained energy dissipation. Since our estimates suggest that it might be important for astrophysical tidal dissipation, it is important to determine how this varies with the amplitude of the tidal deformation and the spin of the planet, which will shall attempt in the next section. The dependence of our results on $\nu_{\alpha},\eta_{\alpha}$ are deferred to \S \ref{Standard}.

\section{Varying $\epsilon$ and $\Omega$}
\label{varepsom}

We now turn to the most important task of this paper: to determine how the efficiency of the turbulent dissipation varies with $\epsilon$ and $\Omega$. The qualitative temporal evolution for a given $\epsilon$ and $\Omega$ is very similar to that outlined in the previous section, so we primarily concentrate here on studying the variation of the turbulent dissipation, since that it what is astrophysically relevant. 

\begin{figure}
  \begin{center}
   \subfigure{\includegraphics[trim=0cm 0cm 0cm 0cm, clip=true,width=0.5\textwidth]{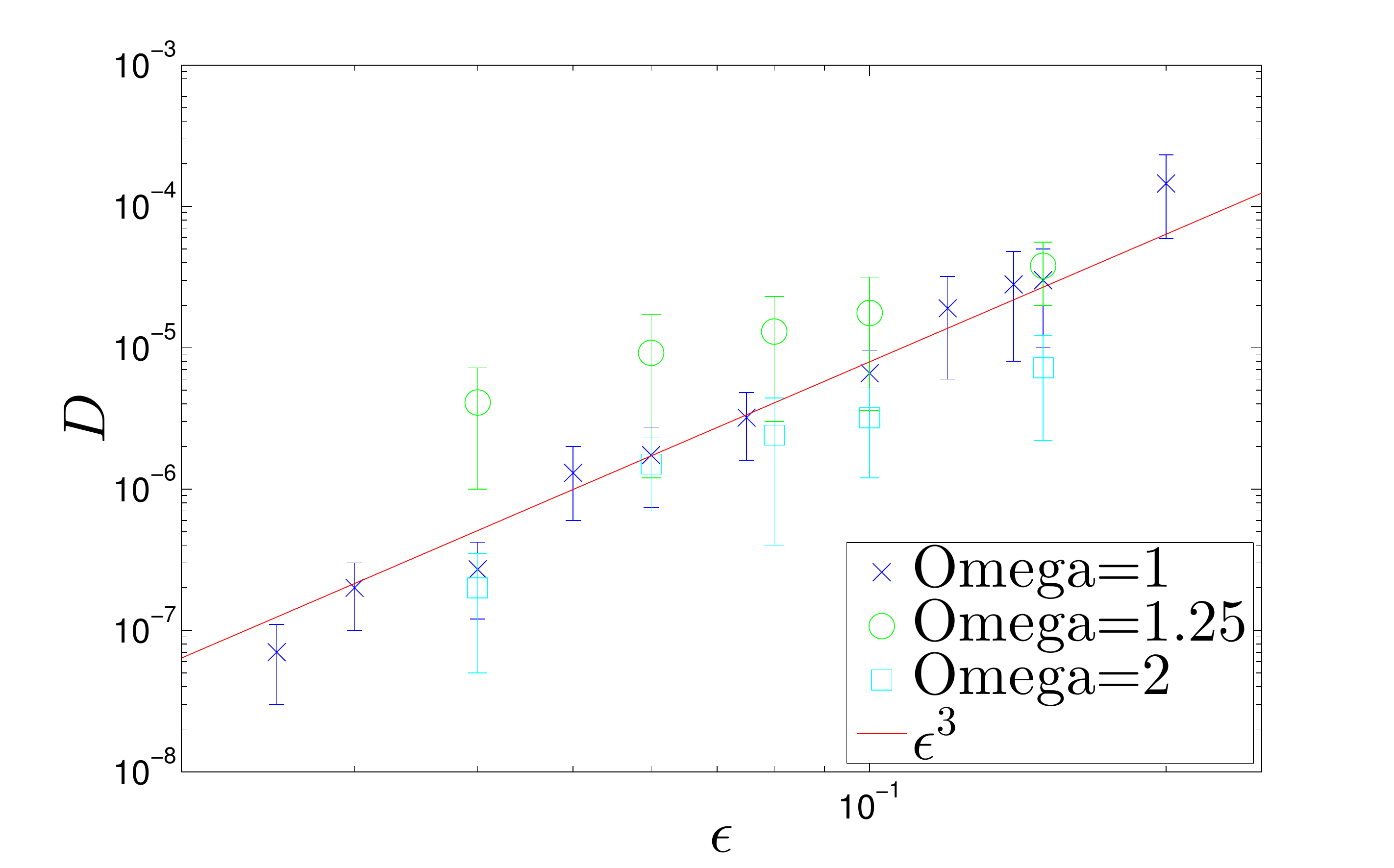} } 
        \end{center}
  \caption{Temporally averaged total turbulent dissipation as $\epsilon$ is varied, from simulations with three different values of $\Omega$. The red line represents $D\propto \epsilon^{3}$, which is roughly consistent with the data throughout this range.}
  \label{2}
\end{figure}

\begin{figure}
  \begin{center}
   \subfigure{\includegraphics[trim=0cm 0cm 0cm 0cm, clip=true,width=0.5\textwidth]{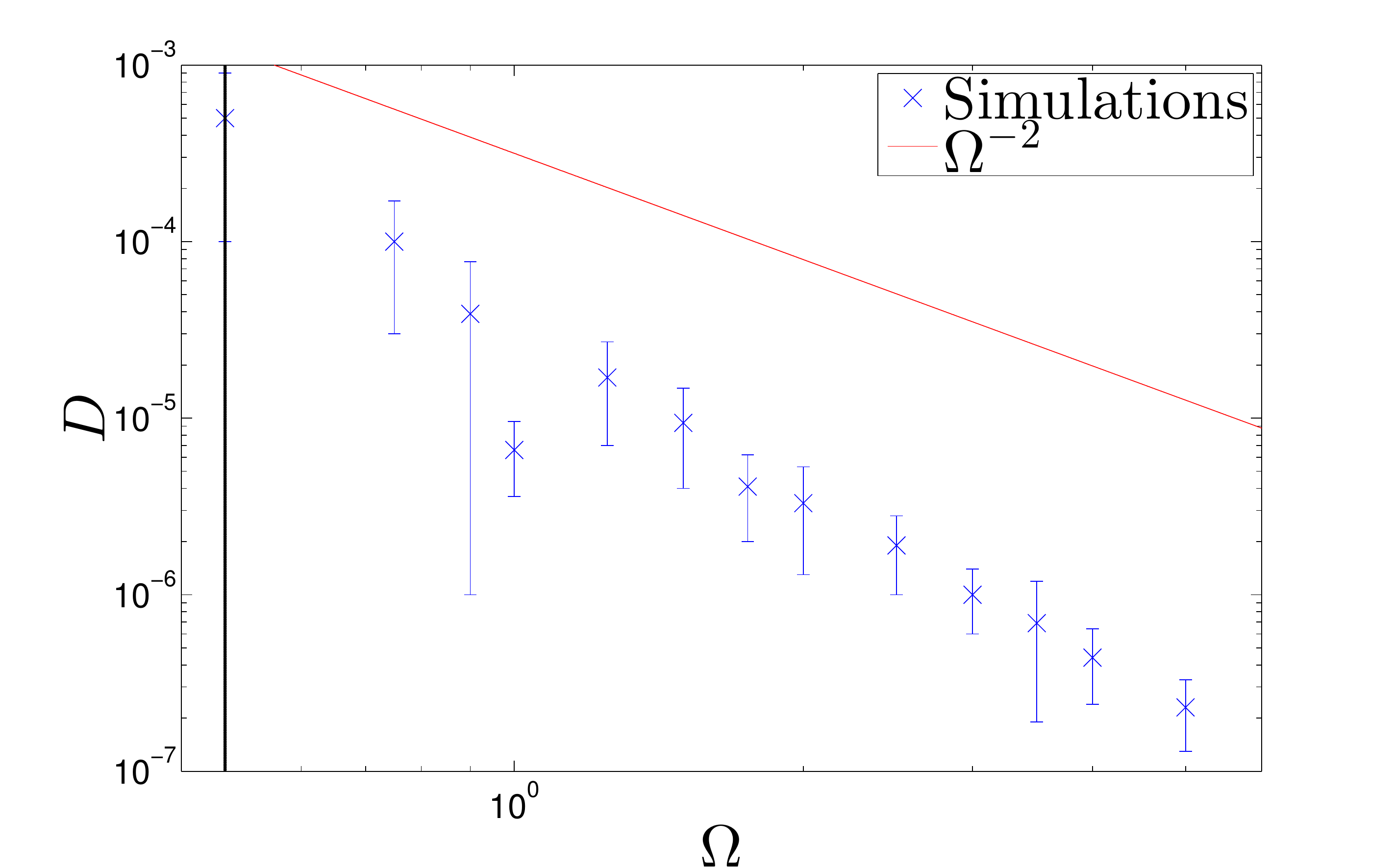} } 
       \end{center}
  \caption{Temporally averaged total turbulent dissipation as $\Omega$ is varied for $\epsilon=0.1$. The red line represents $D\propto \Omega^{-2}$, which is our theoretical prediction. Note that the instability is not present for $\Omega\in [-0.5,0.5)$, which lies to the left of the black line in the figure.}
  \label{3}
\end{figure}

\begin{figure}
  \begin{center}
      \subfigure{\includegraphics[trim=0cm 0cm 0cm 0cm, clip=true,width=0.5\textwidth]{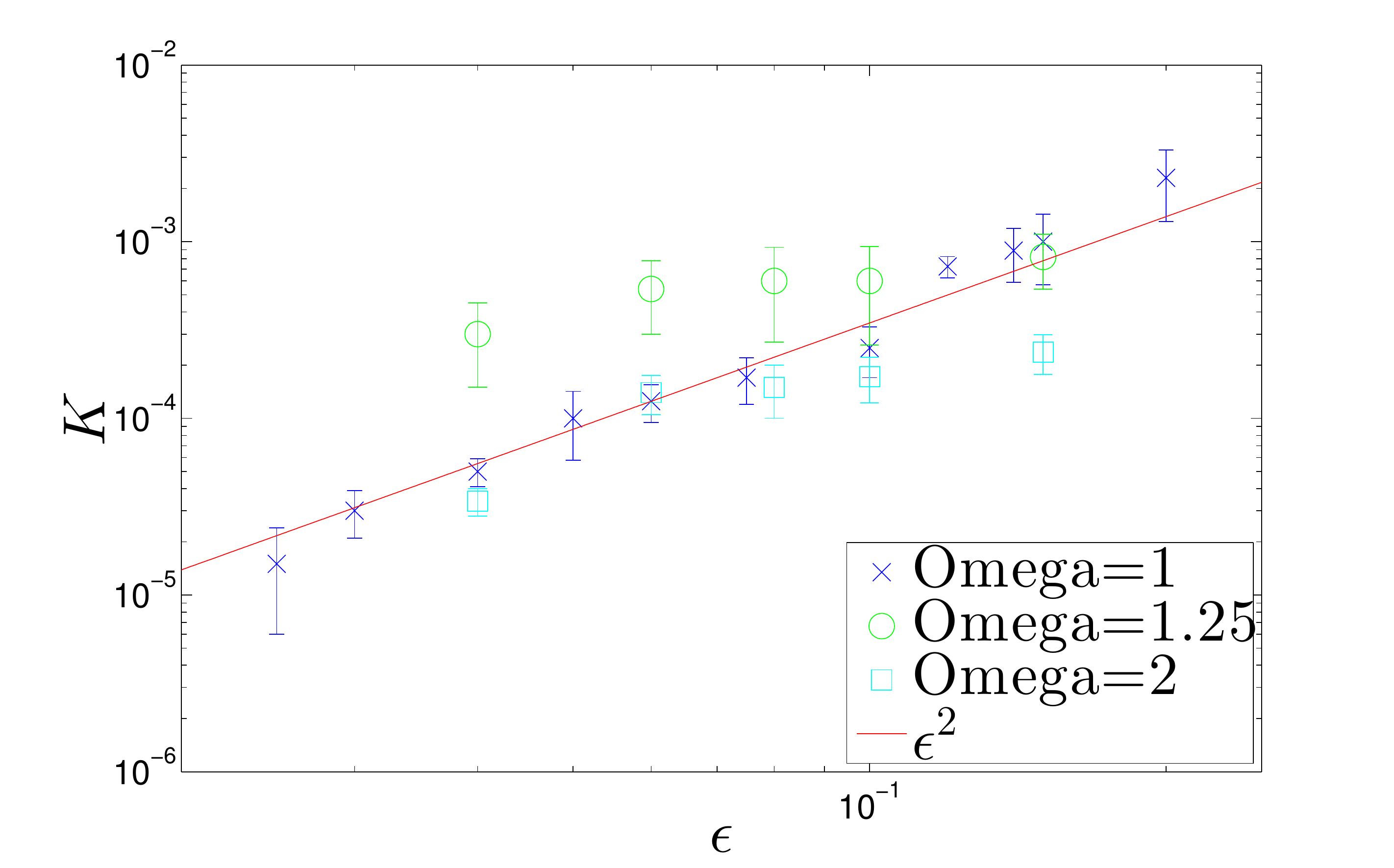} } 
      \end{center}
  \caption{Temporally averaged kinetic energy as $\epsilon$ is varied. The red line shows our prediction 
  that   $K \propto \epsilon^{2}$ i.e. $u \propto \epsilon$.}
  \label{4}
\end{figure}

We calculate a single representative value for the total (viscous plus ohmic) dissipation, as well as the kinetic energy, for a given simulation, by calculating the volume-integrated values and taking a time-average over a time interval of at least $1000$ time units. Since these quantities fluctuate significantly during these simulations, we also indicate the root mean square error of these quantities as error bars, excluding the linear growth phase. 

\begin{table}
\begin{tabular}{l | c | r}
  \hline                        
  $\epsilon$ & $\Omega$ & $k_{peak}$ \\
  \hline
  0.1 & 0.5 & 6.3 \\
  0.1 & 0.9 & 12.6 \\
  0.1 & 1 & 25  \\
  0.1 & 1.25 & 12.6  \\
  0.1 & 2 & 25  \\
  0.1 & 3 & 38 \\
  0.1 & 5 & 63
  \end{tabular}
   \caption{Approximate spherical wavenumber at which the energy per unit logarithmic wavenumber peaks. Note that $k_{peak}=6.3$ corresponds to the largest mode in the box.}
 \label{tablek}
\end{table}

The simple argument outlined in \S \ref{potential} predicts that the dissipation rate (per unit mass) is $D\sim l^{2}\epsilon^{3}\gamma^{3}$, where $l$ is the lengthscale of the modes that dominate the energy. (We retain $\gamma$ here for clarity, though $\gamma=1$ in code units). Fig.~\ref{2} shows results from simulations with various values of $\epsilon$ and $\Omega$. For each $\Omega$, it is roughly true that $D\propto \epsilon^{3}$, which is consistent with our prediction.

Fig.~\ref{3} shows how $D$ depends on $\Omega$ for simulations with $\epsilon=0.1$. Note that the elliptical instability is not excited when $\Omega \in [-0.5,0.5)$ (we have confirmed that initial perturbations simply decay away in this range). In order to predict the dependence of $D$ on $\Omega$ theoretically, we must determine how $l$ depends on $\Omega$. We hypothesise that $l$ is set by the fastest-growing modes, or more specifically, by the wavelength of the fastest-growing modes that fit inside the simulated box. For large $\Omega/\gamma$, the wavevectors of the fastest-growing modes are inclined at $k_{z}/k\approx \gamma/(2\Omega)$ \citep{Craik1989,Kerswell2002}, implying that they have much smaller horizontal than vertical lengthscales. Therefore, the fastest-growing modes have $k_{z}=2\pi/L$, so that $l\sim 1/k\sim L \gamma/\Omega\sim 1/\Omega$, where the last expression is in code units. Therefore, we predict that $D\sim \epsilon^{3}/\Omega^{2}$. Table \ref{tablek} shows the wavenumbers of the energy-dominating modes in the simulations. It is roughly true that $k_{peak}\sim \Omega$, as predicted. However, our prediction in Fig.~\ref{3} is only qualitatively correct: the simulations are closer to $D\propto \Omega^{-3}$ than the predicted $\Omega^{-2}$. We suspect that the disagreement is caused by insufficient resolution for the high $\Omega$ simulations; i.e. since the energy in the high $\Omega$ simulations is peaked at small lengthscales, one should increase the resolution to properly capture the dynamics. Further exploration of the high $\Omega$ simulations is deferred to future investigations.

In Fig.~\ref{4} we plot the averaged kinetic energy for three different values of $\Omega$ as $\epsilon$ is varied. The red line represents a slope of 2, and this is consistent with the prediction that typical velocities scale with $\epsilon$, which provides further support for our simple theory.

Several of the simulations presented in Fig.~\ref{2} have been compared with results obtained using different values of the diffusion coefficients (and different resolutions), with little change in the dissipative properties of the turbulence. Two exceptions to this are the cases with $\epsilon=0.025,\Omega=1$ and $\epsilon=0.03,\Omega=1$, for which the dissipation increases by $50\%$ for a higher resolution (double the number of grid points) case for which $\nu_{4}=\eta_{4}$ is made smaller by $10^{2}$. Presumably this is because decreasing $\epsilon$ reduces the widths of the resonances (e.g.~\citealt{Kerswell2002}), so they eventually become too small for our discrete numerical grid to be able to capture the instability as $\epsilon\rightarrow 0$. In addition, we have performed a higher resolution (double the number of grid points) simulation with $\nu_{4}=\eta_{4}$ that is smaller by $10^{2}$ than the simulations presented in Fig.~\ref{3} with $\Omega=2$ and $5$, finding the dissipation to be the same in each case (to within $15\%$, which is within the root mean square fluctuations). However, these were all performed in a unit box, which is not optimal for large $\Omega$. As we have discussed, the fastest growing mode has a horizontal wavenumber that increases with $\Omega$, so this eventually approaches the dissipation scales as $\Omega$ is increased, for a fixed box aspect ratio. Taking this into account might change our results at the largest $\Omega$ considered. The precise parameters for all of these simulations are written down in Table~\ref{table}. We return to studying the (in)dependence of our results on the diffusivities for $\epsilon=0.1$ and $\Omega=1$ in \S \ref{Standard}.

To summarise the results of this section, we have studied the variation of the turbulent dissipation as a function of $\epsilon$ and $\Omega$, since this is the astrophysically relevant quantity. Our results are roughly consistent with 
\begin{eqnarray}
\label{scaling}
D\approx 10^{-2}\epsilon^{3}\Omega^{-3}.
\end{eqnarray}
There is some uncertainty in both the amplitude and the exponents, especially at large $\Omega$, since in that case the energy is dominated by modes whose lengthscales are not much larger than the grid scale.
An explanation for the numerical amplitude ($10^{-2}$) is that the dominant mode has $k\sim 25$, which is on a smaller scale than the box. The dissipation should therefore be reduced by a factor $\sim1/25^2$ from an estimate that is made assuming $l=1$. We will later use Eq.~\ref{scaling} to estimate the turbulent dissipation resulting from the elliptical instability, which will be applied in more detail to tidal dissipation in fluid planets and stars in \S \ref{discussion}. 

In the purely hydrodynamical simulations presented in BL13, we found that the turbulent dissipation was efficient and exhibited an approximately $\epsilon^{3}$ scaling when $\epsilon \gtrsim 0.15$, 
but that it became much weaker for smaller $\epsilon$ (and faster rotation rates). For $\epsilon\lesssim 0.15$, the elliptical instability led to the formation of columnar vortices that inhibited further instability. This resulted in much weaker dissipation. Here, our results are qualitatively and quantitatively (to within a factor of 2) consistent with the hydrodynamical simulations in BL13 when $\epsilon \gtrsim 0.15$ (in which columnar vortices did not form). However, in this work we did not see a separate regime at small $\epsilon$, corresponding with the formation of vortices, which we attribute here to the presence of magnetic stresses that destroy them and subsequently inhibit their formation. This results in sustained energy injection with moderately efficient dissipation. We will discuss this in more detail in the next section.

\section{Detailed analysis of our simulations}
In this section we analyse some of our simulations in more detail, highlight the differences with our nonmagnetic simulations in BL13, and study the dependence of our results on the diffusion coefficients.
\subsection{Illustrative simulation}

 \begin{figure}
  \begin{center}
             \subfigure{\includegraphics[trim=0cm 0cm 0cm 0cm, clip=true,width=0.45\textwidth] {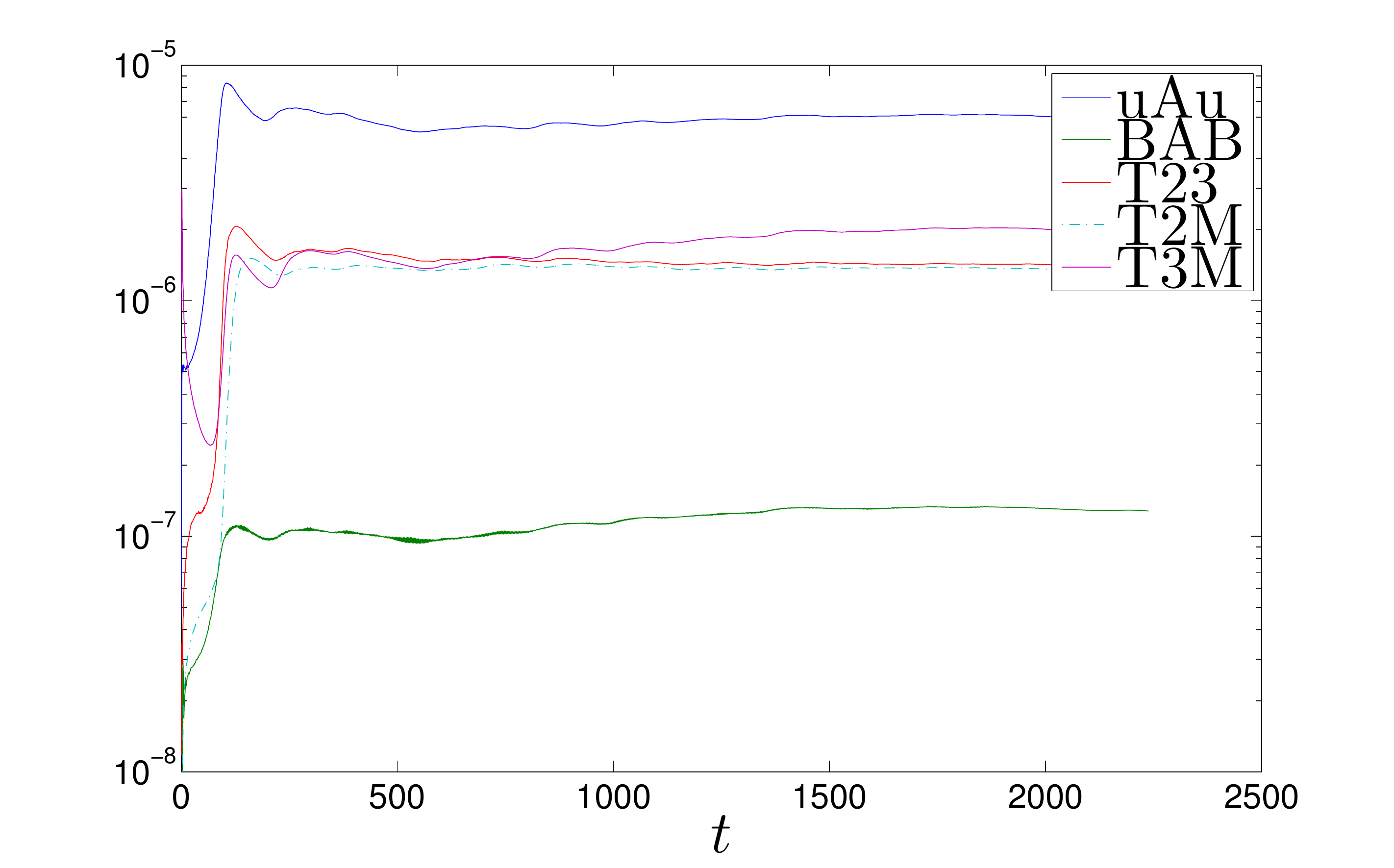} } 
    \end{center}
  \caption{Volume integrated cumulative averages for various quantities in our illustrative simulation with $\epsilon=0.1$, $\alpha=4$ and $\nu_{4}=\eta_{4}=10^{-18}$. The quantities plotted are kinetic energy injection, magnetic energy injection and various nonlinear transfers, as explained in the text.}
  \label{6}
\end{figure}

We first provide a more detailed analysis of the simulation briefly presented in \S \ref{illustrative}. We separate the velocity field into a component with $k_{z}=0$ (vortices) and one with $k_{z}\ne 0$ (waves), as we did in BL13. We denote the $k_{z}=0$ component by the subscript 2D and the $k_{z}\ne 0$ by the subscript 3D. The corresponding volume-integrated quantity measuring the transfer of kinetic energy between the vortices and the waves is
\begin{eqnarray}
T_{23}=\langle\boldsymbol{u}_{2D}\cdot \left(\boldsymbol{u}_{3D}\cdot \nabla\boldsymbol{u}_{3D}\right)\rangle,
\end{eqnarray}
whereas that between the kinetic energy of the vortices and magnetic energy, and that between the waves and magnetic energy, are
\begin{eqnarray}
T_{2M} &=& \langle \boldsymbol{u}_{2D}\cdot \left(\boldsymbol{B}\cdot \nabla\boldsymbol{B}\right)\rangle, \\
T_{3M} &=& \langle \boldsymbol{u}_{3D}\cdot \left(\boldsymbol{B}\cdot \nabla\boldsymbol{B}\right)\rangle,
\end{eqnarray}
where $\langle \boldsymbol{u}\cdot \left(\boldsymbol{B}\cdot \nabla\boldsymbol{B}\right)\rangle=T_{2M}+T_{3M}$. This allows us to quantify the nonlinear transfers from kinetic to magnetic energy and vice versa.

Once the turbulent state is reached after $t\sim200$ (see Fig.~\ref{1}), the flow is primarily dominated by $k_{z}\ne 0$ components (waves), in contrast to the nonmagnetic simulations in BL13, for which the $k_{z}=0$ component (columnar vortices) dominated the kinetic energy. The energy injection into the flow by the elliptical instability is sustained, as can be seen in Fig.~\ref{6}.  Here we plot cumulative averages as a function of time, defined by $\left(\int_{0}^{t}\langle\boldsymbol{u}A\boldsymbol{u} \rangle dt\right) /t$, for example. Reynolds stresses ($\langle\boldsymbol{u}A\boldsymbol{u}\rangle$) dominate Maxwell stresses ($\langle\boldsymbol{B}A\boldsymbol{B}\rangle$), though the latter has nonzero mean, indicating that the elliptical instability is able to inject a very small amount of energy directly into the magnetic field, through the oscillatory stretching of the field by the background flow.

As we described previously, the initial evolution is similar to the nonmagnetic simulations in BL13, in that nonlinear interactions between inertial waves transfer energy into the $k_{z}=0$ component, which tends to generate columnar vortices aligned with the rotation axis. The transfer of energy from waves to vortices is quantified by the quantity $T_{23}$, for which its cumulative average is plotted on Fig.~\ref{6}. This is positive throughout the simulation. However, magnetic stresses extract most of the energy in the vortical component, where this is subsequently cascaded to small scales and dissipated by ohmic diffusion. This is quantified by the observation that $T_{23}\approx T_{2M}$ in Fig.~\ref{6}. Only part of the energy in the wave-like component is transferred to vortices, the rest is either converted to magnetic energy, quantified by $T_{3M}$ (where it is cascaded to small-scales and dissipated by ohmic diffusion), or directly cascaded to small-scales, where it is dissipated by viscosity. The transfer of energy from kinetic to magnetic is due to both the vortical and wave components, with $T_{2M}$ and $T_{3M}$ having similar magnitudes.

\begin{figure}
  \begin{center}
  \subfigure{\includegraphics[trim=0cm 0cm 0cm 0cm, clip=true,width=0.4\textwidth]{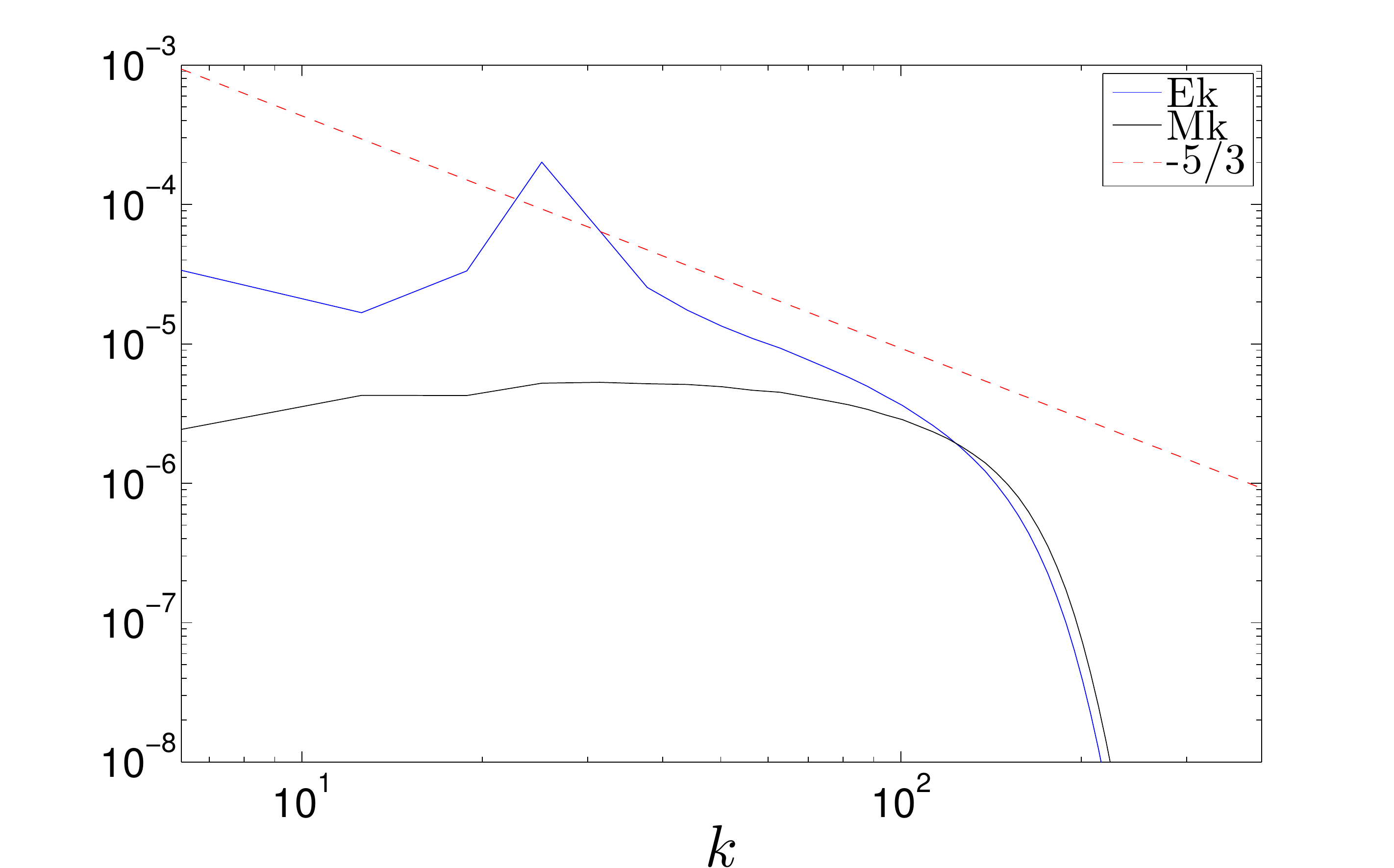} } \\
   \subfigure{\includegraphics[trim=0cm 0cm 0cm 0cm, clip=true,width=0.4\textwidth]{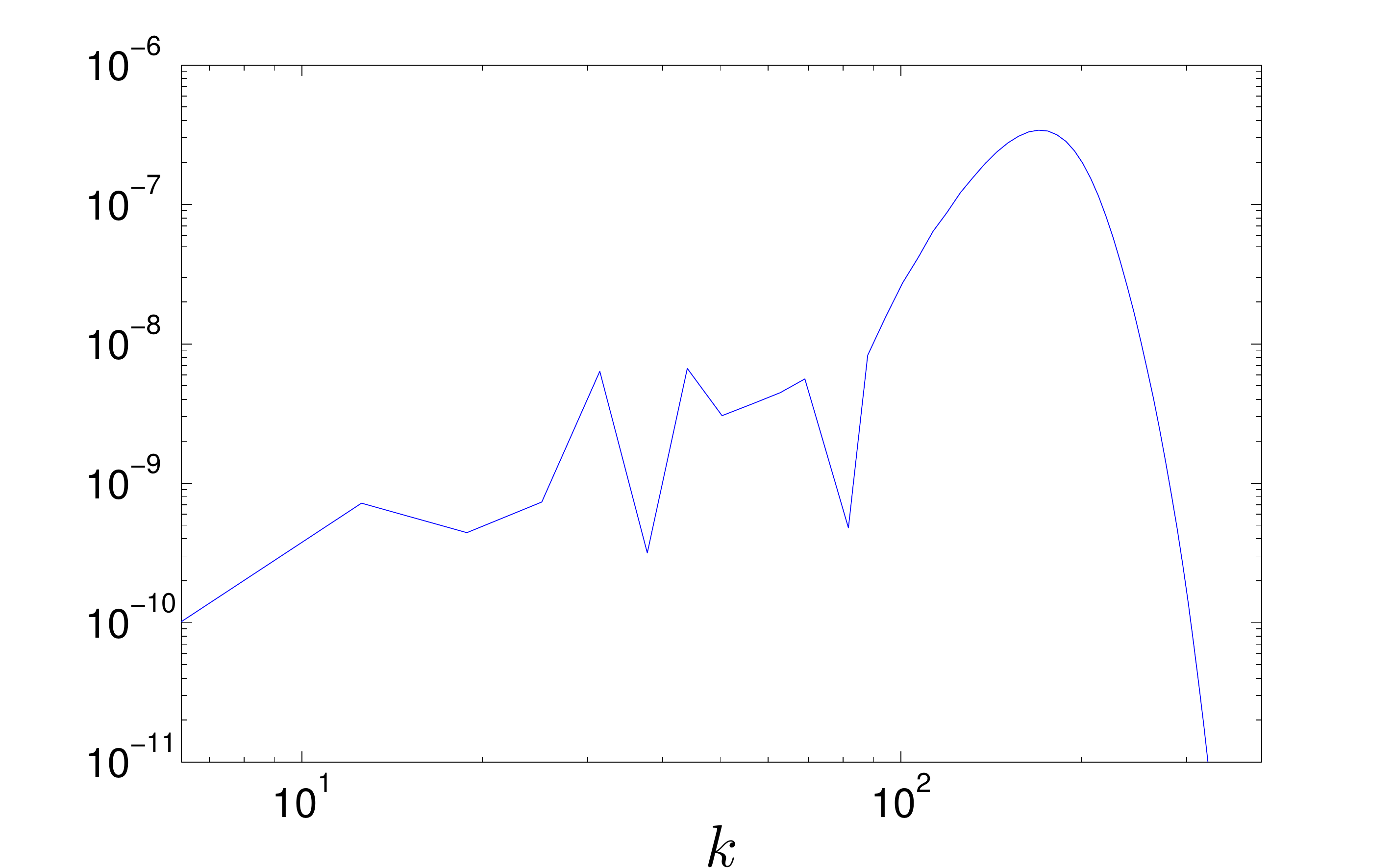} }
   \end{center}
  \caption{Top: Kinetic and Magnetic energy spectra averaged from $t=1000$ to $t=2000$ from our illustrative simulation with $\epsilon=0.1$, $\alpha=4$ and $\nu_{4}=\eta_{4}=10^{-18}$. The red dashed line represents a slope of $-5/3$. Bottom: $|\hat{\boldsymbol{B}}_{k} \cdot \left[\widehat{\nabla \times \left( \boldsymbol{u}\times \boldsymbol{B}\right)}\right]^{*}_{k} + c.c.|$ as a function of $k$, illustrating the scale at which magnetic energy is preferentially generated.} 
  \label{7}
\end{figure}

In the top panel of Fig.~\ref{7}, we plot the spherical shell-averaged kinetic ($E_{k}$) and magnetic energy spectra ($M_{k}$), averaged over the turbulent state from $t=1000$ to $t=2000$. $E_{k}$ (and the energy injection, not plotted) is strongly peaked at $k\approx 25$ for this simulation (however, this varies with $\Omega$, as we discussed in \S \ref{varepsom} and presented in Table \ref{tablek}). The kinetic energy has a spectrum consistent with $-5/3$, corresponding with the Kolmogorov spectrum expected in isotropic and homogeneous hydrodynamical non-rotating turbulence. $M_{k}$ is relatively flat, therefore the magnetic energy per logarithmic bin in $k$ ($\propto k M_{k}$) is strongly peaked at small scales. In the bottom panel of Fig.~\ref{7}, we plot $|\hat{\boldsymbol{B}}_{k} \cdot \left[\widehat{\nabla \times \left( \boldsymbol{u}\times \boldsymbol{B}\right)}\right]^{*}_{k} + c.c.|$. This shows the scale at which magnetic energy is preferentially generated in the flow, which is primarily close to the resistive scale, where $k\sim 200$, which is also the scale at which $E_{k}\sim M_{k}$. This indicates that the flow driven by the elliptical instability acts as a small-scale dynamo.

We have shown in this section that the formation of vortices is inhibited by magnetic stresses. The turbulence represents a competition between the tendency of rapid rotation to produce columnar vortices through nonlinear interactions, and magnetic stresses, which tend to transfer energy from vortices to the magnetic field. The absence of long-lived vortices allows inertial waves to be continually driven, resulting in sustained energy dissipation at a moderately efficient rate, in contrast to the nonmagnetic simulations in BL13.

\subsection{Comparison of different diffusivities, diffusive operators and initial magnetic field configuration}
\label{Pm}
\label{Standard}

\begin{figure}
  \begin{center}
   \subfigure{\includegraphics[trim=0cm 0cm 0cm 0cm, clip=true,width=0.48\textwidth]{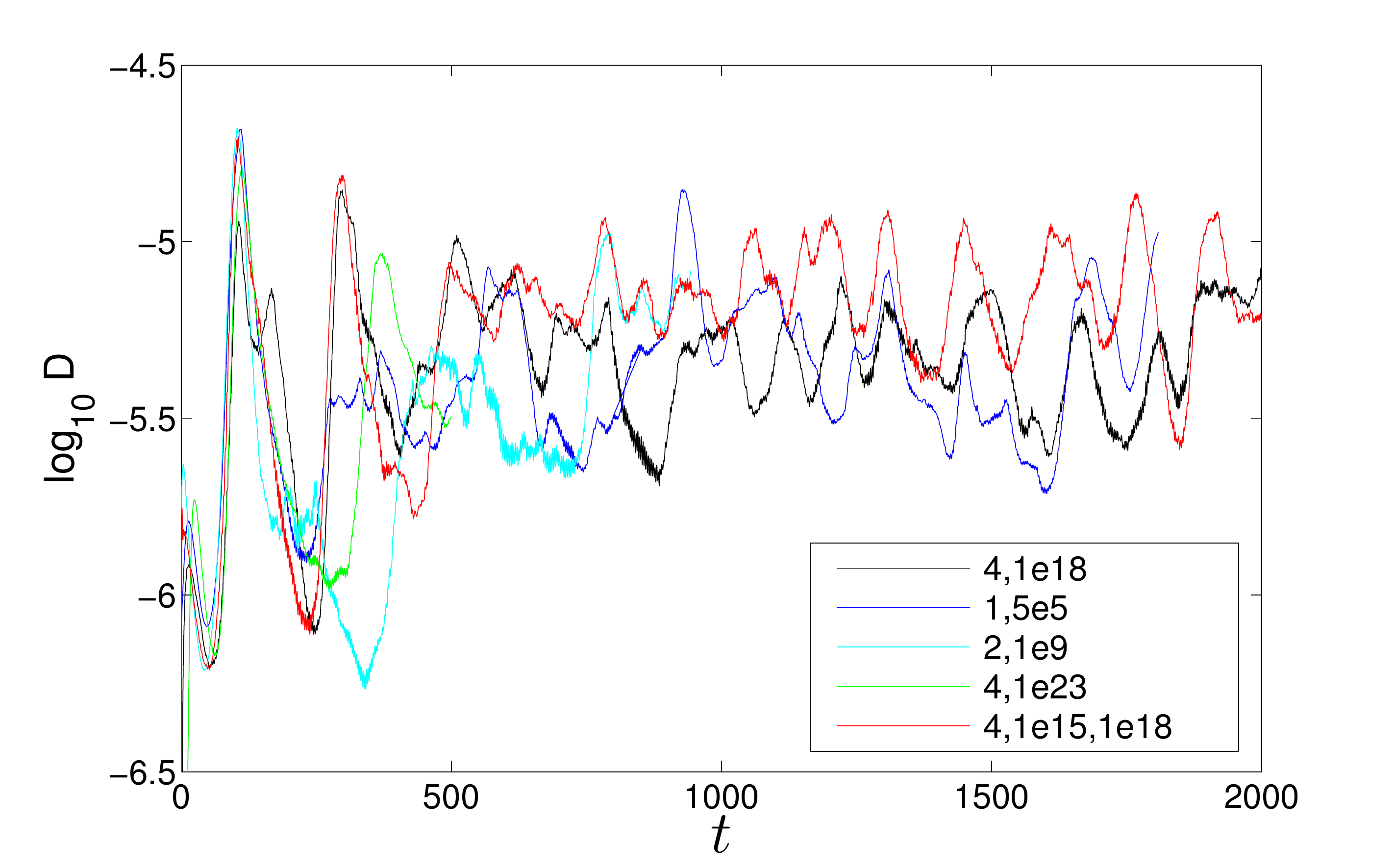} }\\
   \subfigure{\includegraphics[trim=0cm 0cm 0cm 0cm, clip=true,width=0.48\textwidth]{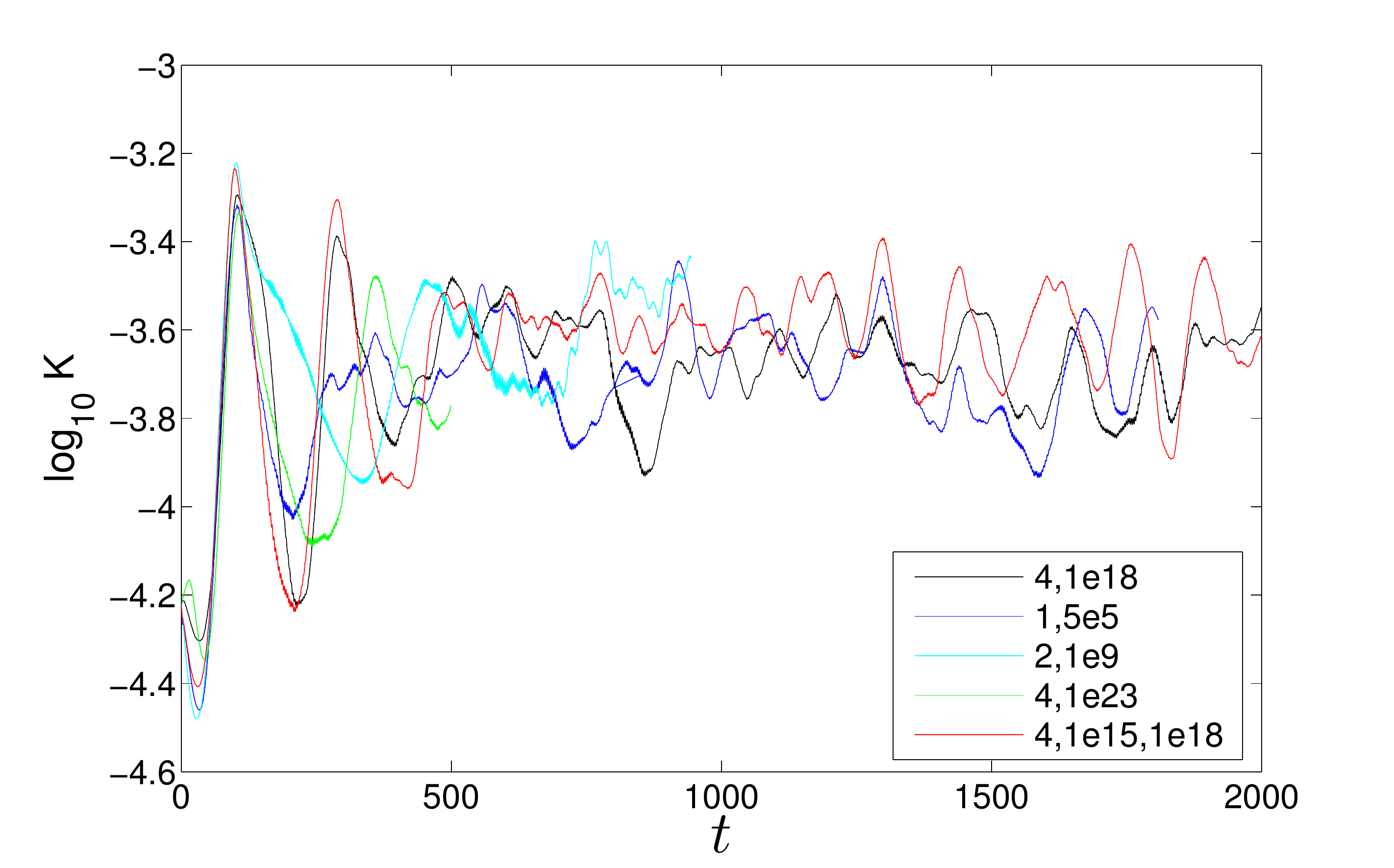} }
    \end{center}
  \caption{Top: Comparison of the total  (kinetic + magnetic) dissipation for various simulations, each with $\epsilon=0.1, \Omega=1$, for various hyper diffusive coefficients and operators. Bottom: Same for kinetic energy. In the legend we label curves by $\alpha,\nu_{\alpha}^{-1}$ and $\eta_{\alpha}^{-1}$ (where the latter differs from the former), respectively.}
  \label{8}
\end{figure}

In this section, we study the dependence of our results on the diffusivities and the diffusion operator, to study the impact that hyperdiffusion has on our results. Since the turbulent dissipation is the most relevant quantity for application to astrophysics, it is essential to determine how robust it is to changes in the diffusive scales, particularly as $\nu_{\alpha},\eta_{\alpha}\rightarrow 0$.

A comparison of various calculations with $\epsilon=0.1$ and $\Omega=1$, is plotted in Fig.~\ref{8}. This presents several simulations with $\alpha=1,2,4$ and various $\nu_{\alpha}$ and $\eta_{\alpha}$ coefficients, as outlined in the legend (and listed in Table~\ref{table}). The simulation with standard viscosity ($\alpha=1$) has $\nu_{1}=\eta_{1}=3\times 10^{-6}$, and is initialised with a magnetic field with strength $B_{0}=10^{-2}$. The main result that can be gleaned from this figure is that neither the kinetic energy of the flow nor the dissipation rate appears to depend on the choice of diffusion operator, or on the values of the (hyper-)diffusion coefficients. This indicates that the form of the dissipative operator appears unimportant in determining the turbulent dissipation, which justifies our use of hyperdiffusion for the majority of the simulations reported in this paper.

\begin{figure}
  \begin{center}
     \subfigure{\includegraphics[trim=0cm 0cm 0cm 0cm, clip=true,width=0.48\textwidth]{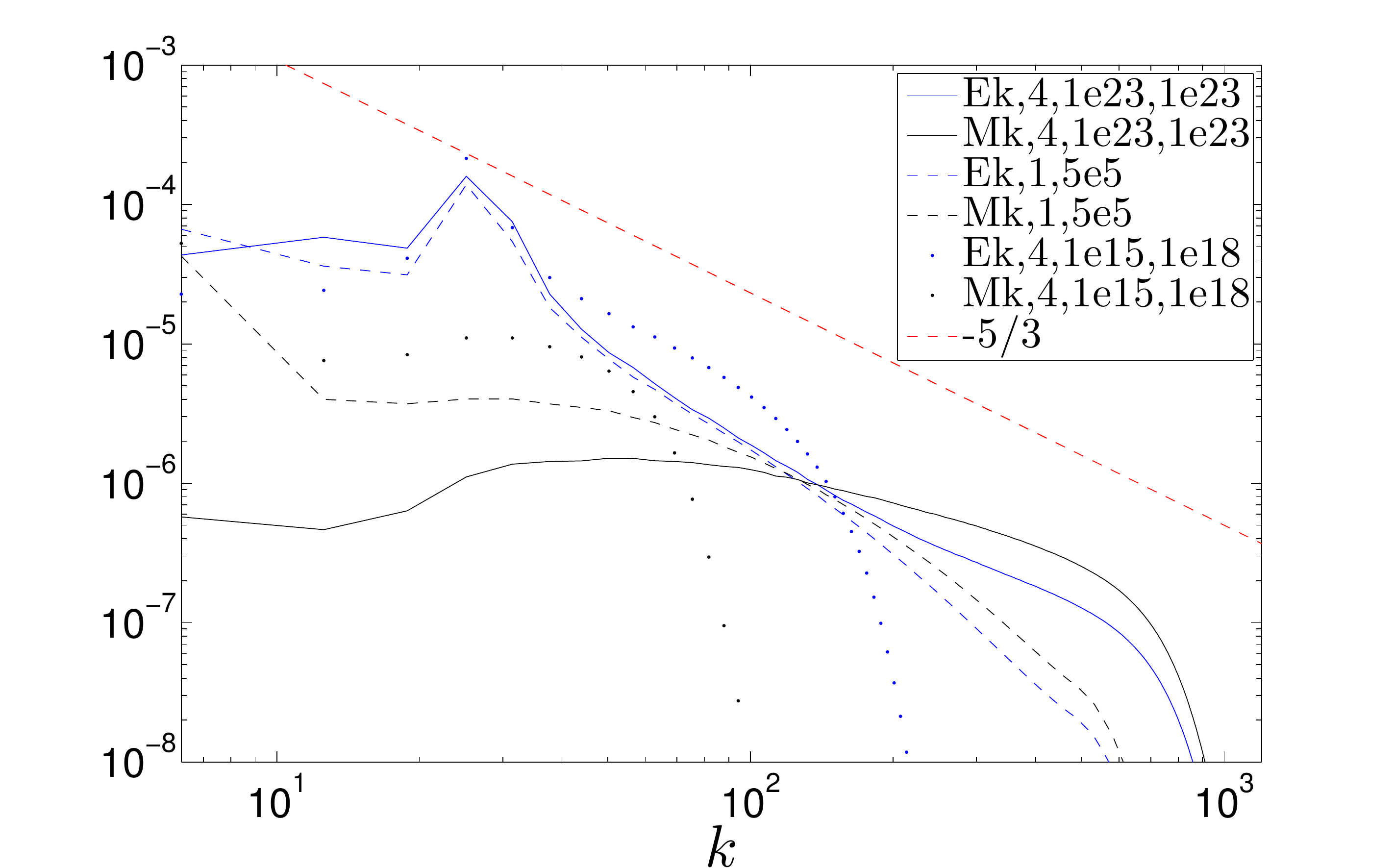} }
   \end{center}
  \caption{Kinetic and Magnetic energy spectra averaged over a time interval of at least 400 (after the linear growth phase) for 
  the same parameters as in Fig.~\ref{7}, but with different diffusion operators and coefficients, as indicated in the legend and described in the text. 
  The kinetic energy is peaked at  large scales, and is relatively independent of the diffusion
  coefficients. By contrast, the magnetic energy is peaked at the smallest scales, and  shifts to higher $k$
  with smaller ohmic (hyper-)diffusivity.}
  \label{9}
\end{figure}

We plot the energy spectra in Fig.~\ref{9} for various simulations with different diffusion coefficients and operators. In this figure, we plot 
 the kinetic (blue lines) and magnetic (black lines) energy spectra for our highest resolution simulation with $\epsilon=0.1$ performed in a $512^3$ box (solid lines), a simulation with standard viscosity performed in a $256^3$ box  (dashed lines), and finally one with an effective Pm $<1$ (dotted lines), with the corresponding diffusion operators and coefficients outlined in the legend. The red dashed line has a slope of $-5/3$. 
 This can be compared with Fig.~\ref{7}, which together illustrate that the peak forcing scale is similar for all simulations with $\epsilon=0.1$ (and $\Omega=1$), at $k\sim 25$, and that there is a forward cascade to small scales with an approximate slope $-5/3$ consistent with the Kolmogorov scaling. The magnetic energy, on the other hand, is strongly peaked at the dissipative scales, since $M_{k}$ is relatively flat in spectral space. 

The parameter regime of relevance for the interiors of stars and giant planets is Pm $=\nu/\eta< 1$. This is a difficult regime to test numerically using $\alpha=1$ diffusion operators, so here we have used hyperdiffusive operators to cover cases with effective values of Pm $< 1$ ($\nu_{\alpha} < \eta_{\alpha}$). An illustration of the energy spectrum for a simulation with $\alpha=4$ and $\nu_{4}=10^{-18},\eta_{4}=10^{-15}$ is presented in Fig.~\ref{9}. Even when the dissipation scale for the magnetic energy is larger than that for kinetic energy, the kinetic spectrum is little changed relative to the fiducial simulation discussed previously (see Fig.~\ref{7}). In addition, the volume-integrated kinetic energy and dissipation rate are also similar (Fig.~\ref{8}). There is no obvious trend as $\eta$ and $\nu$ are varied independently.

Although the range of different $\nu_{\alpha},\eta_{\alpha}$ studied here is necessarily limited, the dissipation in the turbulent state with $\epsilon=0.1$ appears to be well converged, even with the largest diffusion coefficients considered, and does not appear to vary with Pm. The turbulence appears to be somewhat independent of the dissipation scales, which is promising for us to be able to capture the magnitudes of the astrophysically relevant turbulent dissipation in our numerical simulations, which are inevitably limited in their available resolutions.

Finally, we have also run several simulations with various different configurations and magnitudes of $\boldsymbol{B}$. As long as we are in the astrophysically relevant regime, in which $|B_{0}| \ll \epsilon$, the outcome does not depend on the value of $B_{0}$, or on the field configuration (this is no longer true when $|B_{0}| \gtrsim \epsilon$), so these figures are omitted for brevity. Cases we have studied include a vertical field with zero net flux (Eq.~\ref{B0}), a uniform vertical field, and solenoidal white noise perturbations for all components of $\boldsymbol{B}$. This similarity is presumably because the initial magnetic field is weak in comparison with the flow. As along as a magnetic field is present (and ohmic diffusion is sufficiently weak), magnetic stresses are able to significantly modify the turbulence from that observed in the hydrodynamical simulations in BL13. The precise strength and orientation of the initial magnetic field are unimportant.

\section{Dynamo properties of the turbulence}

The elliptical instability has been proposed to drive a dynamo in tidally distorted fluid bodies such as Io's molten core \citep{KerswellMalkus1998}. It has been predicted and experimentally observed that the elliptical instability, when this takes the form of a large-scale ``spin-over" mode, can generate a magnetic field through dynamo action \citep{Lacaze2006,Herreman2010}. Relevant to our configuration, it has been proposed that the elliptical instability could drive a dynamo through a mechanism involving the interaction between linear unstable eigenmodes, which produce a net electromotive force \citep{MizBajMof2012}, even in the absence of ohmic dissipation. However, in cases in which a large-scale coherent flow is not excited, such as the simulations studied in this paper, which may also be relevant for the interiors of fluid planets and stars, it was previously unclear whether flows driven by the elliptical instability could drive a dynamo.

\begin{figure}
  \begin{center}
     \subfigure{\includegraphics[trim=1cm 1cm 1cm 0cm, clip=true,width=0.48\textwidth]{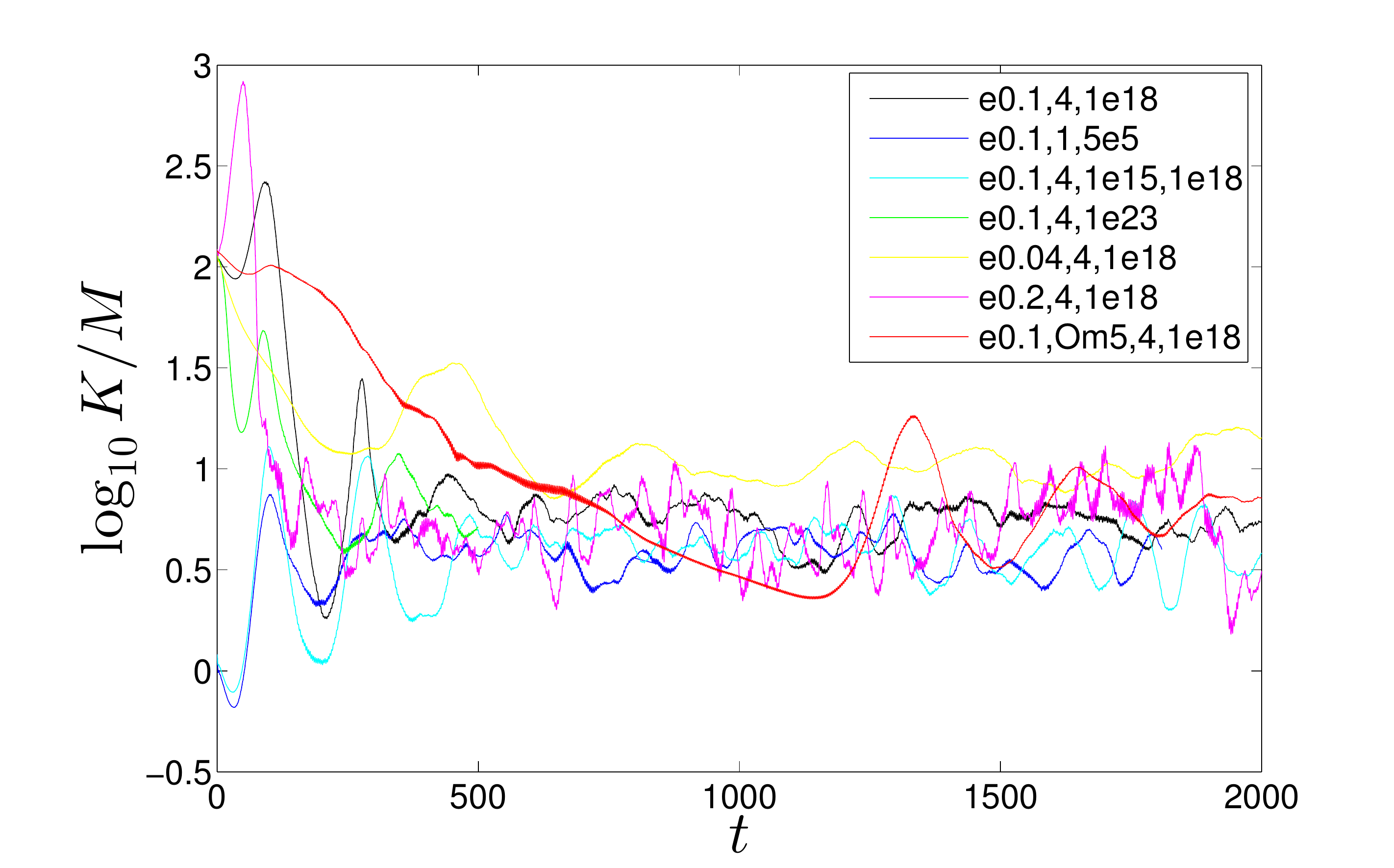} }  
    \end{center}
  \caption{The ratio of kinetic to magnetic energy for a selection of simulations that span the various parameters: $\epsilon$, $\Omega$, diffusivities and initial magnetic field configurations. This shows that
 $K/M\sim 3-10$, and this ratio does not vary significantly throughout our simulations (after the linear growth phase) as various input parameters are varied. We label curves by $\epsilon$, $\Omega$ (only for the last entry, where $\Omega\ne 1$), $\alpha$, $\nu_{\alpha}^{-1}$ and $\eta_{\alpha}^{-1}$, where this differs from the third entry, respectively.}
  \label{3KM}
\end{figure}

\begin{figure}
  \begin{center}
         \subfigure{\includegraphics[trim=0cm 0cm 0cm 0cm, clip=true,width=0.5\textwidth]{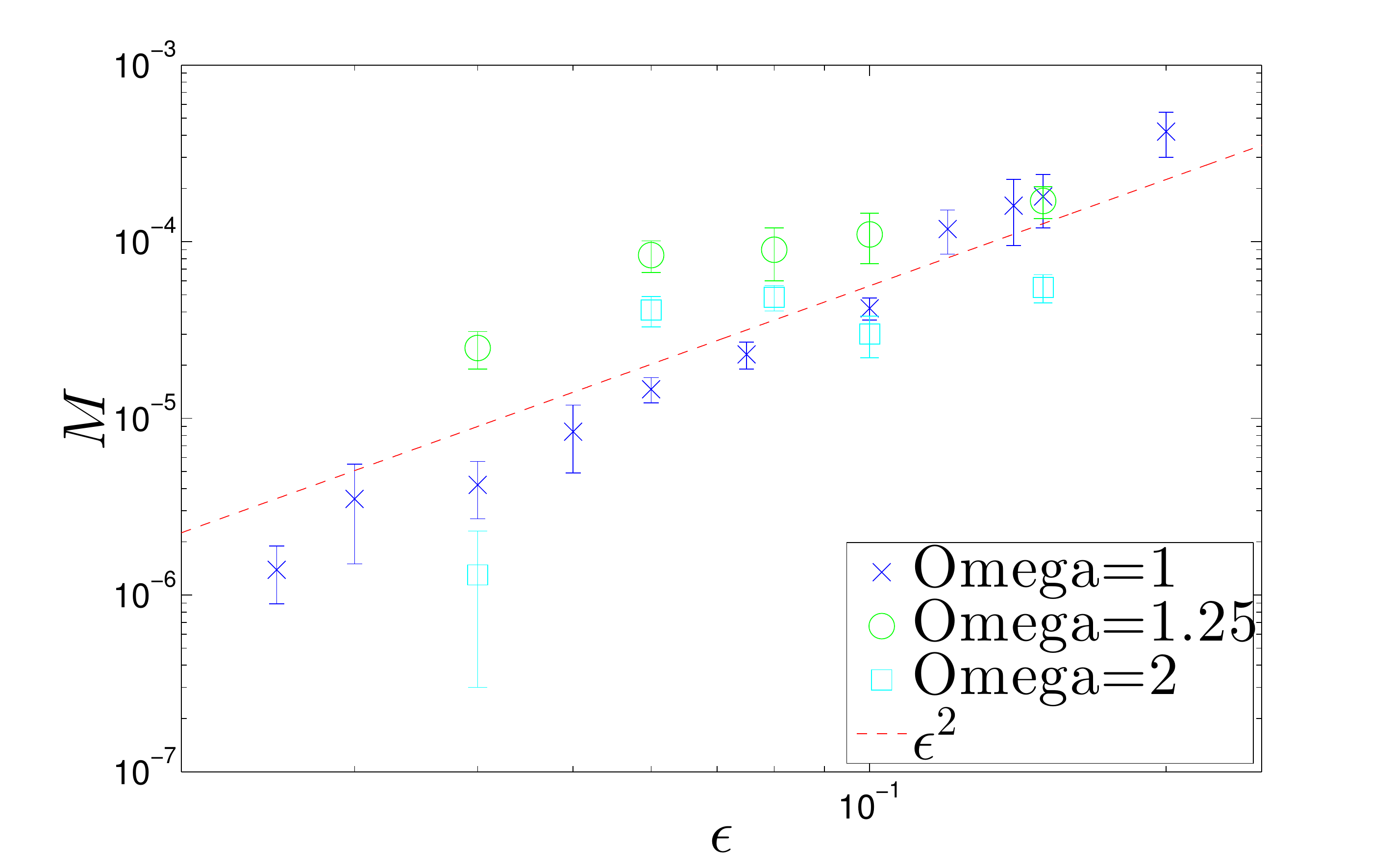} } 
    \end{center}
  \caption{Temporally averaged magnetic energy in various simulations. 
  Comparing with the kinetic energies in these same simulations (Fig. \ref{4}) shows that $M$ is
  always an order-unity fraction of $K$.}
  \label{10}
\end{figure}

As we have observed in Fig.~\ref{7}, the magnetic energy is preferentially generated on the smallest scales, indicating that the elliptical instability acts as a small-scale dynamo. In Fig.~\ref{3KM} we compare the temporal evolution of $K/M$ for various simulations that span the various input parameters, as indicated in the legend. This indicates that the properties of the dynamo are relatively insensitive to our input parameters.
We also plot the temporally and spatially averaged magnetic energy in Fig.~\ref{10} 
for various values of $\epsilon$ and $\Omega$. Comparing with Fig.~\ref{4}, which shows
the kinetic energy in the same simulations, shows that $M$ roughly tracks $K$, and hence $M\propto\epsilon^2$. The magnetic energy, which is initially weak, is amplified and maintained at a value that is a small fraction (typically 0.1-0.3) of the kinetic energy of the flow driven by the elliptical instability.

These results suggest that the elliptical instability is able to generate magnetic fields in tidally distorted planets and stars with energy that is slightly less than the kinetic energy of the turbulent tidal flow driven by the instability. It is unclear whether the elliptical instability is able to generate large-scale magnetic fields, since the simulations reported in this paper only drive a small-scale dynamo. However, this may be the result of the fact that magnetic helicity can evolve only on the slow resistive timescale in our problem, with periodic boundary conditions. Hence, a large-scale dynamo may take a resistive time to develop \citep{Brandenburg2009,Brandenburg2011}. In this case, the use of hyperdiffusion could effectively eliminate the possibility of observing the generation of large-scale magnetic fields. It would be of interest to study the elliptical instability in more detail, in particular by performing global simulations in more realistic geometry, to determine whether a tidally driven large-scale dynamo is possible.

\section{The astrophysical importance of the elliptical instability}
\label{discussion}
Our aim is to determine the magnitude of the turbulent dissipation resulting from elliptical instability, and to assess its importance for tidal evolution. In \S \ref{varepsom} (in particular Figs.~\ref{2} \& \ref{3}), we presented the most important results of our investigation, which are the variation of the turbulent dissipation with the strength of the tidal deformation, as well as the rotation rate of the body. We can use these results to illustrate the astrophysical importance of the elliptical instability. For an estimate of the dissipation resulting from this local model, we take $D=\chi \epsilon^{3}$, where $\chi=10^{-2}$ (we neglect the $\Omega$ dependence in Eq.~\ref{scaling} for the sake of these estimates).

The tidal quality factor that results from dissipation of this form (i.e., from the energy dissipation rate $D=\chi \epsilon^3m_p(2R_p)^2(2\pi/P)^3$, taking $\gamma =2\pi/P$ for these estimates,
after reinserting dimensional parameters) is (BL13) 
\begin{eqnarray}
Q^{\prime} &\approx& \chi^{-1} \left(\frac{m_{p}+m_{\star}}{m_{\star}}\right)\left(\frac{P}{P_{dyn}}\right)^{4} \\
&\approx&  \frac{5 \times 10^{5}}{(\chi/10^{-2})}\left(\frac{m_{p}+m_{\star}}{m_{\star}}\right)\left(\frac{P}{1\; \mathrm{d}}\right)^{4},
\end{eqnarray}
where the relevant quantities were previously defined in \S \ref{potential}.
To put this value in context, the corresponding timescale over which an initially eccentric hot Jupiter is circularised, is (e.g.~\citealt{GoldSot1966})
\begin{eqnarray}
\nonumber
\tau_{e} &=&\frac{4}{63}\frac{Q^{\prime}}{2\pi} \frac{m_{p}}{m_{\star}}\left(\frac{m_{p}+m_{\star}}{m_{p}}\right)^{\frac{5}{3}} \frac{P^{\frac{13}{3}}}{P_{dyn}^{\frac{10}{3}}}
\\  &\approx& 1 \; \textrm{Gyr} \left(\frac{10^{-2}}{\chi}\right) \left(\frac{P}{2.1\; \mathrm{d}}\right)^{\frac{25}{3}},
\end{eqnarray}
where we have taken $m_{p}=10^{-3}m_{\star}$ in the last expression. Similarly, the timescale to synchronise its spin is
\begin{eqnarray}
\nonumber
\tau_{\Omega} &=& \frac{4}{9}\frac{Q^{\prime} r_{g}^{2}}{2\pi} \left(\frac{m_{p}+m_{\star}}{m_{\star}}\right)^{2} \frac{P^{4}}{P_{rot} P_{dyn}^{2}} \\ &\approx&
 1 \; \textrm{Gyr} \left(\frac{10^{-2}}{\chi}\right)\left(\frac{0.5\; \mathrm{d}}{P_{rot}}\right) \left(\frac{P}{5.5 \; \mathrm{d}}\right)^{8},
\end{eqnarray}
where $r^{2}_{g}=0.254$ is the squared dimensionless radius of gyration for Jupiter, and we have taken $m_{p}=10^{-3}m_{\star}$ in the last expression.

These estimates indicate that the elliptical instability is able to circularise the orbit of a hot Jupiter if it began its life with an initially eccentric orbit, as long as its orbital period is shorter than approximately two days. In addition, the elliptical instability appears able to synchronise the spin of a hot Jupiter with its orbit, as long as its orbital period is shorter than about $5.5$ days. This mechanism might therefore contribute to tidal circularisation and synchronisation for the shortest-period hot Jupiters. Note that these timescales are a strong function of the radius of the body, coming through the dependence on $P_{dyn}$, so the instability may be somewhat more efficient in inflated planets, or planets with masses smaller than Jupiter's mass. However, the tidal timescales increase rapidly with $P$, which suggests that the elliptical instability is unlikely to explain the predominantly circular orbits of hot Jupiters out to about 10 days. The strong orbital period dependence of these evolutionary timescales, which is stronger than the case with a constant $Q^{\prime}$, partly arises because this is a nonlinear mechanism of tidal dissipation (a linear mechanism would predict $D\propto \epsilon^{2}$ instead of $\epsilon^{3}$). If we define $P_{e}$ to be the circularisation period, which is the maximum orbital period out to which its orbit can be circularised within $1$ Gyr, and similarly define the synchronisation period $P_{\Omega}$, then $P_{e}\propto \chi^{\frac{3}{25}}$, and $P_{\Omega}\propto \chi^{\frac{1}{8}}$. The dependence of these estimates on $\chi$ is relatively weak as long as the dissipation scales approximately as $\epsilon^{3}$, which is consistent with the simulations reported in this paper (see Fig.~\ref{2}).

The dissipation resulting from the elliptical instability might be thought to play a role in heating the convective interior, preventing contraction and playing a role in inflating the planet (e.g.~\citealt{Cebron2012}). However, the cooling timescale of a Jupiter-mass planet is very short, O(10Myr) (e.g.~\citealt{GuillotShowman2002}).
 If the tidal evolutionary processes occur on a shorter timescale than the age of the system, this heating is transient, and the inflated radius cannot be explained by such a tidal mechanism.
 
The elliptical instability could at best play a modest role in circularising and synchronising the shortest period binary stars. We can estimate the corresponding tidal timescales for two equal mass solar-type close binary stars, to find $P_{e}\approx 2.9$ d and $P_{\Omega}\approx 4.6$ d. Therefore, the elliptical instability could have circularised some of the shortest period close binary stars, and synchronised their spins. However, it does not appear to be sufficiently strong to explain the observed circularisation periods (which was first suggested as a possibility by \citealt{Rieutord2004}), where we would require these processes to operate out to ten days in a few Gyr (e.g.~\citealt{Mazeh2008}, and references therein).

The main uncertainties in applying our results to astrophysics are that our model is a local one, and we have not taken into account any variation in background properties throughout the planet, such as the density, or the influence of realistic boundary conditions, or the possible presence of a solid inner core. In addition, we have omitted buoyancy forces, and the competing influence of turbulent convection. Future global simulations are probably required to study the influence of these additional effects.

Our numerical simulations model tidal synchronisation rather than circularisation.  
For the circularisation problem, the background flow differs from Eq.~\ref{eq:A} in that the amplitude of the tidal deformation also becomes time-dependent (e.g.~\cite{KerswellMalkus1998}; In addition, for non-synchronised bodies, other frequency components are introduced). 
We suspect that our simulations with $\Omega\sim 1$ (i.e.~a tidal frequency equal to $n=\Omega$, for a synchronised body) are reasonable proxies for the circularisation problem
because the linear growth rates in the two cases are similar \citep{KerswellMalkus1998}.  That is why we used
$\chi\approx 10^{-2}$ in our estimate for the circularisation time.  Nonetheless, one should perform 
simulations of the circularisation case to verify our suspicion.  We leave that to future work.

For the synchronisation problem, we have scaled to $\chi\approx 10^{-2}$, as is appropriate
for the $\Omega\sim 1$ simulations in Fig. \ref{3}.  In physical units, those simulations have $\Omega\sim |\Omega-n|$, i.e., $\Omega\gg n$.  Therefore they are only applicable for the initial stages of tidal synchronisation.
For the later stages, when $\Omega\sim n$, our simulations imply that the synchronisation time will 
become much longer.

The estimates of this section indicate that turbulent dissipation resulting from the elliptical instability, in the presence of weak magnetic fields, appears to be of moderate astrophysical importance for tidal dissipation in the interiors of the shortest-period hot Jupiters. However, there are uncertainties in the application of these results directly to astrophysical observations. 

\section{Conclusions}

We have presented the results from a set of magnetohydrodynamical simulations designed to study the nonlinear evolution of the elliptical instability in a small patch of a tidally deformed gaseous planet or star. Our aim was to determine its relevance for tidal dissipation. We continued our investigation, begun in BL13, by adding the additional effect of a weak magnetic field, to determine how this modifies the resulting flow. 

The properties of the turbulence driven by the elliptical instability are qualitatively different in the presence of a weak magnetic field, at least in the astrophysical regime of small ellipticity. In the absence of magnetic fields, the flow becomes organised into columnar vortices aligned with the rotation axis, which we previously observed in our nonmagnetic simulations (BL13). These inhibit the driving mechanism of the instability, significantly reducing the resulting dissipation. But in the  simulations presented in this paper, magnetic stresses prevent the flow from being organised into coherent vortices, resulting in sustained turbulence. 
 We estimated the resulting dissipation by equating the linear growth rate with the nonlinear cascade rate, 
finding $D\sim \epsilon^3$ in nondimensionalised units.  Our simulations roughly confirm that scaling (Fig.~\ref{2}), and further show that the proportionality constant is $D/\epsilon^3\approx 10^{-2}$ (when $\Omega\sim 1$).
Our simulations also show that the elliptical instability acts as a small-scale dynamo, amplifying an initially weak seed magnetic field.

Our main result is that the turbulent dissipation resulting from the elliptical instability, in the presence of a weak magnetic field, is of moderate astrophysical importance. The dissipation appears sufficient to circularise hot Jupiters with $P\lesssim 2.5 $ d, and synchronise their spins with their orbits if $P\lesssim 6 $ d. This mechanism could have also played a role in the circularisation and synchronisation of close binary stars. However, the elliptical instability is unlikely to explain the tidal circularisation of hot Jupiters with longer orbital periods. An as-yet undetermined mechanism might be required to explain circularisation out to such wide orbital separations.

We have neglected a number of effects in our study, such as
 the influence of a more realistic geometry, a solid core, and turbulent convection.
If these could enhance the dissipation by a few orders of magnitude, they could revive the elliptical instability as a more generally applicable mechanism of tidal dissipation.

\section*{acknowledgments}
YL acknowledges the support of NSF grant AST-1109776.

\appendix

\onecolumn
\section[]{Table of simulations}
\label{table}

\begin{tabular}{l | c | c | c | c | c | c | r}
  \hline                        
  $\epsilon$ & $\Omega$ & $B_{0}$ & $\alpha$ & $\nu_{\alpha}$ & $\eta_{\alpha}$ & $N_{x}=N_{y}=N_{z}$ & Comments \\
  \hline
   0.1 & 1 & $10^{-3}$ & 4 & $10^{-18}$ & $10^{-18}$ & 128 & Illustrative case \\
  0.1 & 1 & $10^{-2}$ & 1 & $10^{-5.5}$ & $10^{-5.5}$ & 256 & Standard viscosity \\
  0.1 & 1 & $10^{-3}$ & 2 & $10^{-9}$ & $10^{-9}$ & 128 &  \\
  0.1 & 1 & $10^{-3}$ & 4 & $10^{-23}$ & $10^{-23}$ & 512 &  \\
  0.025 & 1 & $10^{-3}$ & 4 & $10^{-18}$ & $10^{-18}$ & 128 &  \\
  0.025 & 1 & $10^{-3}$ & 4 & $10^{-20}$ & $10^{-20}$ & 256 &  \\
  0.03 & 1 & $10^{-3}$ & 4 & $10^{-18}$ & $10^{-18}$ & 128 &  \\
  0.03 & 1 & $10^{-3}$ & 4 & $10^{-20}$ & $10^{-20}$ & 256 &  \\
  0.04 & 1 & $10^{-3}$ & 4 & $10^{-18}$ & $10^{-18}$ & 128 &  \\
  0.05 & 1 & $10^{-3}$ & 4 & $10^{-18}$ & $10^{-18}$ & 128 &  \\
  0.06 & 1 & $10^{-3}$ & 4 & $10^{-18}$ & $10^{-18}$ & 128 &  \\
  0.075 & 1 & $10^{-3}$ & 4 & $10^{-18}$ & $10^{-18}$ & 128 &  \\
  0.12 & 1 & $10^{-3}$ & 4 & $10^{-18}$ & $10^{-18}$ & 128 &  \\
  0.14 & 1 & $10^{-3}$ & 4 & $10^{-18}$ & $10^{-18}$ & 128 &  \\
  0.15 & 1 & $10^{-3}$ & 4 & $10^{-18}$ & $10^{-18}$ & 128 &  \\
  0.2 & 1 & $10^{-3}$ & 4 & $10^{-18}$ & $10^{-18}$ & 128 &  \\
  0.1 & 1 & $10^{-3}$ & 4 & $10^{-18}$ & $10^{-15}$ & 128 &  \\
  0.1 & 1 & $10^{-3}$ & 4 & $10^{-18}$ & $10^{-17}$ & 128 &  \\
  0.1 & 0.5 & $10^{-3}$ & 4 & $10^{-18}$ & $10^{-18}$ & 128 &  \\
  0.1 & 0.75 & $10^{-3}$ & 4 & $10^{-18}$ & $10^{-18}$ & 128 &  \\
  0.1 & 0.9 & $10^{-3}$ & 4 & $10^{-18}$ & $10^{-18}$ & 128 &  \\
  0.1 & 1.25 & $10^{-3}$ & 4 & $10^{-18}$ & $10^{-18}$ & 128 &  \\
  0.1 & 1.5 & $10^{-3}$ & 4 & $10^{-18}$ & $10^{-18}$ & 128 &  \\
  0.1 & 1.75 & $10^{-3}$ & 4 & $10^{-18}$ & $10^{-18}$ & 128 &  \\
  0.1 & 2 & $10^{-3}$ & 4 & $10^{-18}$ & $10^{-18}$ & 128 &  \\
  0.1 & 2 & $10^{-3}$ & 4 & $10^{-20}$ & $10^{-20}$ & 256 &  \\
  0.1 & 2.5 & $10^{-3}$ & 4 & $10^{-18}$ & $10^{-18}$ & 128 &  \\
  0.1 & 3 & $10^{-3}$ & 4 & $10^{-18}$ & $10^{-18}$ & 128 &  \\
  0.1 & 3.5 & $10^{-3}$ & 4 & $10^{-18}$ & $10^{-18}$ & 128 &  \\
  0.1 & 4 & $10^{-3}$ & 4 & $10^{-18}$ & $10^{-18}$ & 128 &  \\
  0.1 & 5 & $10^{-3}$ & 4 & $10^{-18}$ & $10^{-18}$ & 128 &  \\
  0.1 & 5 & $10^{-3}$ & 4 & $10^{-20}$ & $10^{-20}$ & 256 &  \\
  0.04 & 1.25 & $10^{-3}$ & 4 & $10^{-18}$ & $10^{-18}$ & 128 &  \\
  0.06 & 1.25 & $10^{-3}$ & 4 & $10^{-18}$ & $10^{-18}$ & 128 &  \\
  0.08 & 1.25 & $10^{-3}$ & 4 & $10^{-18}$ & $10^{-18}$ & 128 &  \\
  0.15 & 1.25 & $10^{-3}$ & 4 & $10^{-18}$ & $10^{-18}$ & 128 &  \\
  0.04 & 1.75 & $10^{-3}$ & 4 & $10^{-18}$ & $10^{-18}$ & 128 &  \\
  0.06 & 1.75 & $10^{-3}$ & 4 & $10^{-18}$ & $10^{-18}$ & 128 &  \\
  0.08 & 1.75 & $10^{-3}$ & 4 & $10^{-18}$ & $10^{-18}$ & 128 &  \\
  0.15 & 1.75 & $10^{-3}$ & 4 & $10^{-18}$ & $10^{-18}$ & 128 &  \\
  0.1 & 1 & $10^{-5}$ & 4 & $10^{-18}$ & $10^{-18}$ & 128 & Uniform $B_{z}$ IC \\
  0.1 & 1 & $10^{-5}$ & 4 & $10^{-18}$ & $10^{-18}$ & 128 & Uniform $B_{x}$ IC \\
  0.1 & 1 & $10^{-5}$ & 4 & $10^{-18}$ & $10^{-18}$ & 128 & White-noise for $\boldsymbol{B}$ IC \\
\end{tabular}

\setlength{\bibsep}{0pt}
\bibliography{tidbib}
\bibliographystyle{mn2e}
\end{document}